\documentclass[10pt,a4paper,reqno]{amsart}
\usepackage{amsthm,amssymb,amsmath}
\usepackage{mathrsfs,setspace,pstricks,multicol,multirow, latexsym}
\usepackage{epsfig, subcaption, graphics,graphicx}
\usepackage{dsfont,float}
\usepackage{hyperref}
\hypersetup{
	colorlinks,
	citecolor=blue,
	filecolor=black,
	linkcolor=red,
	urlcolor=black
}
\advance\oddsidemargin by -1.6cm
\advance\evensidemargin by -1.6cm

\numberwithin{equation}{section}

\newcommand\new[1]{}

\newtheorem{Theorem}{Theorem}[section]
\newtheorem{Definition}[Theorem]{Definition}

\newtheorem{Lemma}[Theorem]{Lemma}

\newtheorem{Remark}[Theorem]{Remark}

\newcommand{\noi}{\noindent}

\newcommand{\ds}{\displaystyle}
\newcommand{\RR}{\mathbb{R}}

\newcommand{\la}{\lambda}

\newcommand{\eps}{\varepsilon}

\newcommand{\sig}{\sigma}
\newcommand{\tab}{\hspace*{0.3in}}

\newcommand{\vp}{\varepsilon}

\newcommand{\Sum}{\displaystyle\sum\limits}
\let\emptyset\varnothing

\allowdisplaybreaks

\date{}

\begin{document}
	\pagenumbering{arabic}
	
	\author{Milan Kumar Das}
	\address{Institute of Statistical Science, Academia Sinica, Taiwan}
	\email{das.milan2@gmail.com} 
	
	\author{Anindya Goswami}
	\address{IISER Pune, India}
	\email{anindya@iiserpune.ac.in}
	
	\author{Sharan Rajani}
	\address{Orange Quant Research LLP, Pune, India}
	\email{rajanisharan@outlook.com}
	
	\title[Inference of Binary Regime Model]{Inference of Binary Regime Models with Jump Discontinuities}\thanks{The first author's research was supported by the Ministry of Science and Technology (MOST
108-2811-M-001-628, MOST 108-2118-M-001-003-MY2) of the Republic of China. The research of the second author was supported in part by the SERB MATRICS (MTR/2017/000543), DST FIST (SR/FST/MSI-105), NBHM 02011/1/2019/NBHM(RP)R\&D-II/585, and DST/INT/DAAD/P-12/2020. We also acknowledge National Supercomputing Mission (NSM) for providing computing resources of ‘PARAM Brahma’ at IISER Pune, which is implemented by C-DAC and supported by the Ministry of Electronics and Information Technology (MeitY) and Department of Science and Technology (DST), Government of India.}
	
	\addtocounter{footnote}{-1} \vskip 1 true cm
	
\begin{abstract}
Identifying the instances of jumps in a discrete-time-series sample of a jump diffusion model is a challenging task. We have developed a novel statistical technique for jump detection and volatility estimation in a return time series data using a threshold method. The consistency of the volatility estimator has been obtained. Since we have derived the threshold and the volatility estimator simultaneously by solving an implicit equation, we have obtained unprecedented  accuracy across a wide range of parameter values. Using this method, the increments attributed to jumps have been removed from a large collection of historical data of Indian sectorial indices. Subsequently, we have tested the presence of regime-switching dynamics in the volatility coefficient using a new discriminating statistic. The statistic has been shown to be sensitive to the transition kernel of the regime-switching model. We perform the testing using Bootstrap method and find a clear indication of presence of multiple regimes of volatility in the data. A link to all Python codes is given in the conclusion. The methodology is suitable for analyzing high frequency data and may be applied for algorithmic trading.
\end{abstract}
\maketitle
\noi {\bf Key words:} Regime switching models, Jump models, Statistical inference, Threshold method\\
	{\bf AMSC:} 91G70, 60J76, 62F12, 62M09
\section{Introduction}
Following the seminal work of Black and Scholes \cite{BS73}, the geometric Brownian motion (GBM) was adopted by several authors and market practitioners for modelling the risky asset price dynamics. The scope of this model was extended in subsequent literature in various different directions, many of which include possibility of jump discontinuities in the model. Merton \cite{Mer76} was the pioneer to introduce such models in the 1976's to price an option. Since then various aspects of jump diffusion models (see \cite{Con03}, \cite{Ell90}, \cite{Mad98} and references therein) are being discovered in numerous studies. Needless to say, the jump models are now widely accepted and being used by market practitioners. The main reason behind the popularity of jump models is that  the models with continuous paths cannot explain the empirically observed occasional sudden big changes in simple return series.

The regime switching models are another class of models, getting attention in financial literature after the influential work of Hamilton \cite{hamilton}. Such models allow certain simplistic variability of market parameters. The market parameters are modelled as pure jump processes with finitely many states which correspond to Market regimes. However, these states are not directly observable in the financial market. We refer the readers to \cite{BAS}, \cite{BO}, \cite{BU}, \cite{MAT}, \cite{FAN}, \cite{AGMKG}, \cite{AJP}, \cite{AS}, \cite{JA}, \cite{LI}, \cite{LIA}, \cite{PA} for more details. The models for asset dynamics which incorporate both the jumps and regime switching have also been investigated in many works including \cite{Cos}, \cite{MAN}, \cite{Ell07}, \cite{GMR19} and references therein. However, an unified  inference of jump diffusion models with regime switching is still missing in the literature.

In this paper, we address the inference problem for a class of binary regime-switching jump diffusion models. In \cite{DG19}, the statistical inference of a class of binary regime switching models of financial time series data has been addressed assuming absence of jump discontinuities. In that a particular type of test statistics has been used which is suitable for a model where the low volatility regime occurs with a low probability. A theoretical extension of that study for the counter part, i.e., models where the high volatility regime occurs with low probability seems straightforward. However, there is a practical challenge. The real time series data of financial asset prices do exhibit jump discontinuities. Around the time of jump, the historical volatility appears too high. Such high volatility occurrences cannot be explained using just a regime-switching extension of geometric Brownian motion model. Because, after all, theoretically the switching GBM has continuous path almost surely. In other words, sudden large changes in return series should be attributed to jump discontinuities. Therefore, the jumps have to be identified and removed before inferring the high regime in the regime-switching volatility dynamics.

Several techniques have been  developed to infer the parameters of jump models which include the maximum likelihood approach \cite{ait, IY2021}, the Markov chain Monte Carlo approach \cite{era}, and threshold estimators \cite{AG2020, jac08, man08}.
Unlike the constant threshold approach presented in \cite{man08}, we propose a new threshold technique to disentangle the jump from the continuous part.
In the constant threshold method, depending on the time granularity of the data, firstly a threshold value of return is fixed and up-crossing of that are classified as the jump times. Following the removal of detected jumps from the return series, the volatility coefficient is estimated. It is also known that if the threshold does not depend on the diffusion coefficient, the jump identification becomes poorer for larger value of the coefficient. This is so, as some rare but increasingly more large increment in Brownian motion gets misclassified as jumps (see  \cite{FM2019}, \cite{FL2020}, \cite{LM}). In \cite{LM}, the returns are normalized with spot volatility estimates and the threshold is obtained from the outliers of these values. In \cite{FM2019} proposes an optimal threshold which is obtained by minimizing the mean squared error in parameter estimation. This leads to an iterative method. A different type of iteration has been used in \cite{FL2020}. On the other hand we have reduced the false positive error in a non-iterative manner by computing the threshold and the volatility estimator simultaneously by solving an implicit equation in single variable. A numerical scheme for solving that equation is also presented.

As par our knowledge, the approach of the present paper is absent in the literature. We call the resulting estimator as maximal, because this maximizes the contribution of diffusion term in explaining return series and diminishes the false-positive error. In addition to this, in maximal threshold, one can set a confidence level for controlling the false positive error. Although controlling that has been the main focus, the error in volatility parameter estimation is also controlled as a consequence.

Following the removal of jumps, we model the derived continuous part of the data using a binary regime switching GBM model. The regime switching dynamics is a pure jump process. That could be either a finite state continuous time Markov chain or a semi-Markov chain and is used for modelling volatility dynamics. The main difference between a Markov chain and a semi-Markov chain is in its instantaneous transition rate. It is just a constant matrix for the homogeneous Markov chain whereas for the semi-Markov chain it is a matrix valued function of the sojourn time. So far the applicability is concerned, the SMGBM models are superior to the Markovian counterpart for its greater flexibility in fitting the inter transition times. Since such flexibility in model fitting directly leads to improvements in derivative pricing, the statistical comparison between the Markov and semi-Markov regime switching samples becomes particularly important. For the comparison purpose, we propose a discriminating statistic. Its sampling distribution desirably varies drastically, with varying choices of instantaneous rate parameter in the regime switching dynamics.

The discriminating statistic is constructed using some descriptive statistics of squeeze and expansion duration of Bollinger band, which seems to be the most natural approach. The sampling distributions of the descriptive statistics of these durations under a particular model hypothesis do not have nice forms. Hence the inference cannot be done analytically. In spite of this limitation, one can surely obtain an empirical distribution of the statistics using a reliable simulation procedure. This is a standard approach and is termed as the typical realization surrogate data approach in Theiler et al \cite{JT}. Readers may find application of this approach in many other texts including but not restricted to \cite{SS1} and \cite{JD}.

This paper is organised in several sections and subsections. The detection of jumps is developed in Section \ref{section2}. In section \ref{duration}, we give details of obtaining squeeze and expansion durations. These durations are used to construct the discriminating statistic in section \ref{section3}. Here, we also explain the rejection procedure of any composite null hypothesis based on this statistic. Section \ref{Simulation} contains the discretization of some important class of regime switching models. These discretizations are the key steps for sampling from the null hypothesis. In section \ref{empirical} we apply the inference technique to some empirical data. We end this paper with some concluding remarks in Section \ref{conclusion}.

\section{Inference of jumps}\label{section2}
\noi In this section, we present a statistical method for detection of jump discontinuities in the asset return data using a threshold approach. Since both, the diffusion noise and the jump factor contribute to the second order moment of the return process, a mere knowledge of return's variance is not adequate to solve the calibration problem of both jump and diffusion coefficients. Finer information can be extracted by classifying the returns using a threshold. Volatility coefficient, the annualized coefficient of diffusion term, is estimated in the fixed threshold approach, after inferring the jump coefficients. In the proposed approach of maximal threshold, the threshold value and the volatility coefficients are obtained simultaneously.
\subsection{Model hypothesis} Following the indirect approach \cite{GMR}, for detection of jump  discontinuities, we first consider the following simplified auxiliary continuous-time model of asset price process $S:=\{S_t\}_{t\in [0,T]}$, given by
	    \begin{equation}
	        dS_t = \mu S_{t-}dt + \beta S_{t-}dW_t + S_{t-}dM_t \label{EmpMod}
	    \end{equation}
with  $S_0>0$, where $W=\{W_t\}_{t\in [0,T]}$ is the standard Brownian motion, and  $M=\{M_t\}_{t\in [0,T]}$ is a compound Poisson process. In particular, $M$ is given by  $M_t = \sum_{i=1}^{\mathcal{N}_t} \xi(i)$, in which $\mathcal{N}=\{\mathcal{N}_t\}_{t\in [0,T]}$ is a Poisson process with intensity $\Lambda$ and  $\xi:=\{\xi(i)\}_{i=1,2, \ldots}$ is a sequence of independent random variables with identical cumulative distribution function (cdf) $F$ having mean zero and a finite variance. We assume that $W$, $\mathcal{N}$ and $\xi$ are independent to each other.  We also note that \eqref{EmpMod} implies the following model
\begin{align}\label{ret}
\frac{S(i) - S(i-1)}{S(i-1)} -\mu \Delta = \beta \sqrt{\Delta}Z(i) + (M_{i\Delta}-M_{(i-1)\Delta})
\end{align}
for the discrete time series $(S(0), S(1), S(2),\ldots, S(N))$ having time step $\Delta$ (in year unit), where  $\{Z(i)\}$ are independent, identically distributed (i.i.d.) standard normal random variables. Here $S(i)$ stands for $S_{i\Delta}$ in \eqref{EmpMod}.
From a given equi-spaced data the one-step simple return is given by $r(i) = \frac{S(i) - S(i-1)}{S(i-1)}$. The average $\overline{r}$ of $\{r(1), \ldots, r(N)\}$ is given by $\overline{r} = \frac{1}{N}\sum_{i=1}^{N}r(i)$. Using \eqref{ret} and the model assumption that $E\xi(i)=0$, it is clear that $\overline{r}$ is an unbiased estimator of $\mu\Delta$. Furthermore, we assume the following two conditions on $F$
\begin{itemize}
    \item [(A1)] $F(-1)$ is assumed to be zero,
    \item [(A2)] $F(0-)=F(0)$, i.e., zero is a point of continuity of $F$.
\end{itemize}
Assumption (A1) ensures positivity of $S$, whereas (A2) implies that $P(\xi=0)=0$. In other words, (A2) prohibits jumps of size zero in simple return. This does not impose any practical restriction on \eqref{EmpMod}, but clarifies the selection of the cdf of jump size and the jump intensity. We recall that in the classical Merton's Jump Diffusion (MJD) model $F$ is the cdf of one less than lognormal and hence MJD obeys (A1) and (A2). Although \eqref{EmpMod} is more general than classical MJD, we still call \eqref{EmpMod} as the MJD model now onward for the terminological convenience.

\subsection{Motivation of Threshold Method}
\noindent For empirical study via model fitting one needs to estimate all the parameters, namely historical $\mu, \beta, \Lambda$ and $F$ where $F$ turns out to be a functional parameter. Although the jump discontinuities are apparent in a continuous time process, the identification becomes ambiguous when the process is observed in discrete time. We observe in the following simple lemma that the false positive error in jump detection can be lowered as the time step decreases.
\begin{Lemma} \label{lem2.1} Assume $\Lambda =0$. Given any $\hat{p}\in (0,1)$, and $c>0$, there exists a sufficiently small $\Delta>0$, such that $P(\bigcup_{i=1}^N \{|r(i) - \mu \Delta| \ge c\})< \hat{p}$.
\end{Lemma}

\begin{proof} Due to \eqref{ret}, under $\Lambda =0$, $\{r(1), \ldots, r(N)\}$ is a sequence of i.i.d normal variables with mean $\mu \Delta$ and variance $\beta^2 \Delta$. Let $A_i:= \{\omega\in \Omega \mid |r(i) - \mu \Delta| \ge c\}$, then
$$ P(A_i)= 2\left(1-\Phi \left(\frac{c}{\beta\sqrt{\Delta }}\right)\right)$$
for each $i$, where $\Phi$ denotes the cdf of the standard normal distribution. Therefore,
$$P\left(\bigcup_{i=1}^N A_i\right) = 1- P\left(\bigcap_{i=1}^N (\Omega \setminus A_i)\right) = 1- \prod_{i=1}^N (1- P( A_i))= 1- \left(2\Phi \left( \frac{c}{\beta\sqrt{\Delta }} \right)-1 \right)^N.$$
Hence the lemma is true, i.e., left side is less than $\hat{p}$ if and only if
\begin{align}\label{ineq}
\nonumber &2 \Phi \left(\frac{c}{\beta\sqrt{\Delta }}\right) -1 >  (1-\hat{p})^{1/N} = (1-\hat{p})^{\frac{\Delta}{T}}\\
\textrm{or, }&  \frac{T}{\Delta} \ln \left( 2\Phi \left(\frac{c}{\beta\sqrt{\Delta }}\right)-1\right) > \ln (1-\hat{p})
\end{align}
where $T$ denotes the length of the time horizon (i.e., $T = N \Delta$). A direct calculation gives that the left side of the above inequality vanishes as $\Delta$ goes to zero
whereas right side is a fixed negative quantity. Hence there is a sufficiently small $\Delta>0$ such that \eqref{ineq} is true. Hence the proof is completed.
\end{proof}

\noindent Due to the continuity (see (A2)) of $F$ at zero, given any $\hat{p}>0$, there is a positive $c'$ such that for any $x\in (-c',c')$, one has $|F(x)-F(0)|<\hat{p}/2$. This gives
\begin{align}\label{F(c)}
0\le F(c')-F(-c')<\hat{p},
\end{align}
that is $P(|\xi|<c')<\hat{p}$. In other words, if such a value of $c'$ is set as threshold, the chance of false negative in jump detection becomes less than $\hat{p}$.

\begin{Remark}
In view of Lemma \ref{lem2.1} and the implication \eqref{F(c)} of (A2), the noise sources, i.e., $Z$ (diffusion noise) and $M$ (jump noise) in the discrete time series (sampled from \eqref{ret}) can be separated using an appropriate threshold parameter $c$ or $c'$ with confidence $1-\hat{p}$, if the time discretization is sufficiently small. Thus the value of the threshold plays a key role in such classification of noise. While an estimation of $c'$ as in \eqref{F(c)} is practically impossible from a real data, it is still not hard to compute the greatest lower bound so that \eqref{ineq} holds for any value of $c$ higher than that. A threshold, thus obtained, keeps the false positive error bounded by $\hat{p}$. We explain this in the remaining part of this section.
\end{Remark}

\begin{Definition}\label{defi1}
In view of the above remark, if $\hat\beta$ is an estimator of $\beta$,
we compute $\hat{c}:= \gamma \hat{\beta}$ where
\begin{align}\label{c}
\gamma :=\sqrt{\Delta} \Phi^{-1}\left(\frac{1+(1-\hat{p})^{\frac{\Delta}{T}}}{2}\right),
\end{align}
is a known positive constant. Given a time series data, we  say a jump has occurred at $i^{\textrm{th}}$ time step if $|r(i) - \overline{r}|$ is more less than $\hat{c}$ and at that instance, $r(i) - \overline{r}$ gives the value of jump size.
\end{Definition}

\subsection{Maximal Threshold}

\noi We consider an estimator (SD) of the standard deviation of one-step return by \footnote{
This is a consistent estimator. Furthermore, as  $\Delta$ is supposed to be chosen significantly small, for example few minutes in year unit, the magnitude of $r(i)-\bar{r}$ becomes small. On the other hand in a typical time series data $N$ is considerably large. Hence the bias in $SD^2$ is negligible.}
\begin{equation*}
SD^2 = \frac{1}{N}\Sum_{i=1}^{N} \Big(r(i) - \overline{r} \Big)^2.
\end{equation*}
We also note that $SD^2$ does not depend on $\hat{p}$. Finally, as a consequence of \eqref{ret}, an estimator $\hat{\beta}$ of $\beta$, can be chosen to follow
\begin{align}\label{beta}
SD^2 = \hat{\beta}^2\Delta + \hat{\Lambda}\Delta V,
\end{align}
where $\hat{\Lambda}$ and $V$ are estimators of jump intensity and the variance of jump sizes. To see this, we recall the variance formula of compound Poisson process, i.e., $VAR(M_t) = \Lambda t E(\xi^2)$. We set
\begin{align} \label{lambda:est}
\hat{\Lambda} := \frac{card (I_{\hat c})}{T},
\end{align}
where $I_{\hat c}= \{i\in \{1,\ldots, N\} \mid |r(i) - \overline{r}| > {\hat c}\}$ and $card(A)$ denotes the  cardinality of a set $A$. Then $\hat{\Lambda}$ is a plug-in estimator of $\Lambda$ based on the maximum likelihood estimation. Furthermore, $V$ is set as $V := \frac{\sum_{i\in I_{\hat c}}(r(i)-\Bar{r})^2 }{card \left( I_{\hat c}\right)} = \frac{\sum_{i\in I_{\hat c}}(r(i)-\Bar{r})^2
}{\hat{\Lambda}T}$ (using \eqref{lambda:est}).
By substituting the expressions of $SD^2$, $V$ and $\hat{\Lambda}$ in \eqref{beta}, we obtain the following equation
\begin{equation}\label{para:est}
G(\hat{\beta}):= \hat{\beta}^2 -\frac{SD^2}{\Delta} +\frac{1}{T}\sum_{i\in I_{\gamma\hat{\beta}}}(r(i)-\Bar{r})^2=0.
\end{equation}
Using the solution of above, one can evaluate $\hat{\Lambda} = \frac{card\left( I_{\gamma\hat{\beta}}\right)}{T}$ and subsequently $V=\frac{\sum_{i\in I_{\gamma\hat{\beta}}}(r(i)-\Bar{r})^2}{\hat{\Lambda}T}$. Clearly \eqref{para:est} has a trivial solution $\hat{\beta} =0$. This  leads to $\hat{\Lambda}=\frac{N}{T} =1/\Delta$ and $V=SD^2$. However, we are interested in the nontrivial solution where $\hat{\beta} >0$, since larger the $\hat{\beta}$ lesser the time points attributed to jumps. To be more precise, we are looking for the largest solution to \eqref{para:est}, as we wish to minimize the false-positive error in jump detection. We recall that a low false-positive error means the return series is explained by the diffusion term as much as possible.

\begin{Theorem} \label{2.5} The equation \eqref{para:est} has a non-trivial solution for sufficiently small $\hat{p}$.
\end{Theorem}
\begin{proof} We need to prove the existence of a nontrivial zero of $G$ (the function of $\hat{\beta}$ that is defined on the left side of \eqref{para:est}).
It is evident that if $\hat{\beta}$ is more than $SD/\sqrt{\Delta}$, $G(\hat{\beta})$ is strictly positive. Hence  any positive solution, if exists, lies in $(0, SD/\sqrt{\Delta}]$. Again, \eqref{c} implies that $\lim_{\hat{p}\to 0} \gamma =\infty$. Therefore,
\begin{align*}
    \lim_{\hat{p}\to 0} \sum_{i\in I_{\gamma\hat{\beta}}}(r(i)-\Bar{r})^2 = 0
\end{align*}
almost surely as $I_{\gamma\hat{\beta}}$ becomes empty for a sufficiently large $\gamma$. Thus for any $ \hat{\beta} \in (0, SD/\sqrt{\Delta})$, there is a sufficient small $\hat{p}$ such that $G(\hat{\beta})$ is strictly negative. We fix such $\hat{p}$. Since $G$ is positive on $(\frac{SD}{\sqrt{\Delta}},\infty)$, we conclude that $G$ is bounded below and rises(need not be monotonic) from negative to positive as $\hat{\beta}$ increases. This confirms existence of a zero of $G$ if each discontinuity of $G$ is due to a negative jump. Moreover, being bounded below, if $G$ additionally possesses right-continuity and has positive first order derivative at each point of continuity, would permit a minimizer $\hat{\beta}_{\min}$, say.\\

\noindent For establishing above properties of $G$, we consider the term $\frac{1}{T}\sum_{i\in I_{\gamma\hat{\beta}}} (r(i)-\Bar{r})^2$, which clearly is a non-increasing right continuous step function of $\hat{\beta}$. Since $G$ is an addition of this term with a continuous function $\hat{\beta}^2$ and a constant, $G$ should also be right continuous, having only negative jumps and has positive first and second order derivatives at the points of continuity. Therefore $G$ has zeros (at least one) on $(\hat{\beta}_{\min}, SD/\sqrt{\Delta})$.
\end{proof}
\begin{Remark} \label{rem1}
\noindent It is important to note that \eqref{para:est} is to be solved numerically for a given data set. Due to the discontinuity of $G$, uniqueness of non-trivial zero is not obvious. As $G$ is strictly positive beyond $SD/\sqrt{\Delta}$, the largest zero of $G$ is finite and that is strictly positive and can be determined uniquely, provided $G$ is ever negative. Also we identify this as the desired solution to \eqref{para:est}. We also note that $G$ is not only right continuous, has right derivative too. More precisely $G'(\beta)= 2\beta$, where $G'$ denotes the right derivative of $G$. Hence the following Newton-Raphson algorithm with initial point $\frac{SD}{\sqrt{\Delta}}$ converges to the largest solution to \eqref{para:est} rapidly
\begin{align}\label{betan}
\beta_{n+1} = \beta_n -\frac{G(\beta_n)}{2\beta_n}, \quad\forall n\ge 0, \textrm{ and } \quad \beta_0 = SD/\sqrt{\Delta}.
\end{align}
This is evident as $G$ has jumps of negative size and has positive first and second order derivatives at the points of continuity and the initial point is on the right of the desired solution. The Figure \ref{ntsol} illustrates this.

\begin{figure}[h] \includegraphics[width=0.2\linewidth]{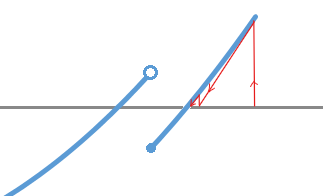} \caption{Convergence of gradient descent to the largest solution of \eqref{para:est}}\label{ntsol} \end{figure}
\end{Remark}

\begin{Definition}[Maximal Estimators]\label{defi}
We denote the largest solution to \eqref{para:est} as $\hat{\beta}_{\max}$ and call that as the \emph{maximal estimator} of $\beta$ parameter. Moreover, we call the product $\gamma \hat{\beta}_{\max}$ as  the maximal threshold and denote that by $\hat{c}_{\max}$.
\end{Definition}
\noindent  While $\hat{\beta}_{\max}$ is defined above, the limit of the iterative sequence \eqref{betan} gives its value, provided it exists. Theorem \ref{2.5} assures existence for sufficiently small $\hat{p}$. Clearly the above mentioned threshold method differs from the methods based on fixed threshold or iterative thresholds. The present approach is termed as the maximal threshold method as the accuracy of jump-detection is optimized by controlling the false-positive error. This is achieved by obtaining the threshold and the volatility  together by solving an implicit equation. The following theorem having an independent theoretical interest, answers to the uniqueness question of \eqref{para:est} on a restricted set and also indicates a location of the estimator.
\begin{Theorem}\label{2.6}
For sufficiently small $\hat{p}$, \eqref{para:est} has a unique solution on $\left(\frac{SD}{\sqrt{2 \Delta}}, \frac{SD}{\sqrt{ \Delta}}\right]$ and no solution on $\left(\frac{SD}{\sqrt{\Delta}}, \infty\right)$.
\end{Theorem}
\begin{proof}
Let $\beta_0= \frac{SD}{\sqrt{2 \Delta}}$ and $\gamma_0:= \max_{i\le N}\left( \frac{|r(i) - \overline{r}|}{\beta_0}\right)$. We fix $\hat{p}$ sufficiently small, so that $\gamma$ in \eqref{c} is more than $\gamma_0$. Hence for all $\hat{\beta}$ not less than $\beta_0$, $\gamma\hat{\beta}$ is greater than  $\max_{i\le N} |r(i) - \overline{r}|$ and consequently $I_{\gamma\hat{\beta}}$ is empty. Therefore, at  $\hat{\beta}= \beta_0$, $G$ is equal to $- {\beta_0}^2$, a negative quantity; and continues to increase continuously on $(\beta_0,\infty)$ and is positive at $\hat{\beta}= \frac{SD}{\sqrt{\Delta}}$. Hence the proof.
\end{proof}
\noindent We have so far discussed about computation of $\hat{\beta}_{\max}$, for the sole reason of obtaining a threshold value $\hat{c}_{\max}$. The threshold enables us to isolate the jumps from diffusion noise in return series with confidence $1-\hat{p}$. The class of data points having jumps, can then separately be analyzed for inference of the jump size distribution $F$. Moreover, the return series driven by only diffusion can also be analyzed for fitting some other sophisticated diffusion models. We elaborate this aspect in subsequent sections.

\subsection{Consistency of \texorpdfstring{$\hat{\beta}$}{TEXT}} Next we discuss the consistency issues of the estimator $\hat{\beta}$ as $\Delta \to 0$. We note that $\hat{\beta}$ depends on $\gamma$ which again depends on both $\Delta$ and the choice of the misclassification probability $\hat{p}$.
For making $\hat{\beta}$ solely dependent on $\Delta$, the probability of mis-classification should be taken as dependent on $\Delta$ and be sent to zero as $\Delta\to 0$ instead of fixing a constant pre-assigned value $\hat{p}$. It is also not difficult to anticipate that the decay rate of the misclassification probability plays a key role in establishing consistency of the estimator. Here we choose
\begin{align} \label{p}
\hat{p}=\min\left(1, \alpha \Delta^{\ln(1/\Delta)}\right)
\end{align}
for some $\alpha >0$. Clearly, $\hat{p}\to 0$ as $\Delta \to 0$. Also any desired value can be assigned to $\hat{p}$ by choosing $\alpha$ accordingly. Now onward, we denote the right side of \eqref{c} as $\gamma(\Delta)$ in which $\hat{p}$ is given by \eqref{p}. For proving that this choice of $ \hat{p}$ gives the desired consistency, the following lemma is needed.

\begin{Lemma}\label{lemma2}
Let $\gamma(\Delta)$ be as above. Then as $\Delta$ tends to $0$,
\begin{enumerate}
    \item $\gamma(\Delta)$ converges to zero and
    \item $\frac{\Delta \ln \frac{1}{\Delta}}{\gamma(\Delta)^2}$ converges to zero.
\end{enumerate}
\end{Lemma}
\begin{proof}
It follows from the formula 7.1.13 of \cite{Abr48} that for any $x\geq 0$
\begin{align}\label{gauss}
\frac{1}{x+\sqrt{x^2+4}}< \sqrt{\frac{\pi}{2}}e^{\frac{x^2}{2}}\left(1-\Phi(x)\right)<\frac{1}{x+\sqrt{x^2+\frac{8}{\pi}}}.
\end{align}
Therefore for any $\eps\ge 0$, we have that
\begin{align}\label{gauss:bnd}
\frac{e^{\eps x^2}}{x+\sqrt{x^2+4}}< \sqrt{\frac{\pi}{2}}e^{(\frac{1}{2}+\eps)x^2}\left(1-\Phi(x)\right)<\frac{e^{\eps x^2}}{x+\sqrt{x^2+\frac{8}{\pi}}}.
\end{align}
Now substituting $x$ by $\frac{x}{\sqrt{\frac{1}{2} + \eps }}$ in the left hand side of the inequality \eqref{gauss:bnd}, we deduce
\begin{align*}
    e^{x^2}\left(1-\Phi\left(\frac{x}{\sqrt{\frac{1}{2} + \eps }}\right)\right)>\sqrt{\frac{2}{\pi}}\frac{\left(\sqrt{\frac{1}{2} + \eps }\right)e^{\frac{\eps}{\frac{1}{2} + \eps } x^2}}{x+\sqrt{x^2+4\eps +2}}.
\end{align*}
Since the right side diverges to infinity as $x\to \infty$, given any $M>0$, there exists a $x_0>0$ such that for all $x\geq x_0$, we have 
$ 1-\Phi\left({x}/{\sqrt{\frac{1}{2} + \eps }}\right)>Me^{-x^2}.$
Again, since $\Phi$ is non-decreasing, we obtain
\begin{align}\label{bnd:1}
  {x}/{\sqrt{\frac{1}{2} + \eps }}<  \Phi^{-1}\left(1-Me^{-x^2}\right).
\end{align}
On the other hand, substitution of $x$ by $\sqrt{2}x$ in the right side of the inequality \eqref{gauss}, gives convergence of $ e^{x^2}\left(1-\Phi\left(\sqrt{2} x\right)\right)$  to $0$ as $x\to\infty$, in other words, given any $M>0$, there exists a $x'_0>0$ such that for all $x\geq x'_0$, we have the following
\begin{align}\label{bnd:2}
  \sqrt{2} x >  \Phi^{-1}\left(1-Me^{-x^2}\right).
\end{align}
With slight abuse of notation, we denote $x_0:=\max\{x_0,x'_0\}$. Then \eqref{bnd:1} and \eqref{bnd:2} hold for any $x\geq x_0$. For a fixed $\hat{p}\in (0,1)$ and sufficiently large $x\geq x_0$, let $y>0$, be such that the following holds
\begin{align*}
   1-Me^{-x^2} = \frac{1+(1-\hat{p})^y}{2}.
\end{align*}
By using the Taylor's series expansion of $(1-\hat{p})^y$, from above we obtain
\begin{align*}
    x^2 = -\ln{\left(\frac{\hat{p}y}{2M}+o(y)\right)}.
\end{align*}
Then there is a $y_0>0$ such that for all $y<y_0$, we have 
\begin{align}\label{bnd:3}
   \sqrt{\ln{\left(\frac{4M}{\hat{p}y}\right)}} > x = \sqrt{-\ln{\left(\frac{\hat{p}y}{2M}+o(y)\right)}} > \sqrt{\ln{\left(\frac{M}{\hat{p}y}\right)}}.
\end{align}
Now by using \eqref{bnd:3} in \eqref{bnd:1}, and by setting $y=\frac{\Delta}{T}$, for all $\Delta<\Delta_0$ with $\Delta_0=Ty_0$, we deduce
\begin{align*}
    \sqrt{\ln{\left(\frac{MT}{\hat{p}\Delta}\right)}}&<\sqrt{\frac{1}{2} + \eps }\Phi^{-1}\left(\frac{1+(1-\hat{p})^{\frac{\Delta}{T}}}{2}\right) =\left(\frac{1}{2} + \eps \right)^{1/2} \frac{\gamma( \Delta)}{\sqrt{\Delta}},
\end{align*}
provided $\hat{p}=\min\left(1, \alpha \Delta^{\ln(1/\Delta)}\right)$. Without loss of generality, assume $\Delta_0$ sufficiently small so that $\hat{p}<1$ for any $\Delta<\Delta_0$. By substituting $M=1$ in the above inequality, and by squaring both sides we get
\begin{align}\label{bnd:4}
    0<\ln\left(\frac{T}{\alpha }\right)+ \ln \left(\frac{1}{\Delta}^{\ln(1/\Delta)}\right)+\ln\left(\frac{1}{\Delta}\right) < \left(\frac{1}{2} + \eps \right) \frac{\gamma^2(\Delta)}{\Delta},~~ \forall~ \Delta< \Delta_0.
\end{align}
Or,
\begin{align*}
  \frac{\Delta}{\gamma^2( \Delta)} \ln(T/\alpha) + \frac{\Delta}{\gamma^2( \Delta)}\left(\ln(1/\Delta)\right)^{2}+\frac{\Delta}{\gamma^2( \Delta)}\ln(1/\Delta)<  \left(\frac{1}{2} + \eps \right).
\end{align*}
Note that the second term of the left hand side of the above inequality is strictly dominating over the other terms as $\Delta\to 0$ and also the sum is bounded. This implies that the terms other than the second
vanish as $\Delta \to 0$. Thus we prove our second claim, i.e., $\frac{\Delta \ln \frac{1}{\Delta}}{\gamma(\Delta)^2} \to 0$ as $\Delta \to 0$. Again, by using \eqref{bnd:3} in \eqref{bnd:2}, and by setting $y=\frac{\Delta}{T}$, for all $\Delta < \Delta_0$, we get that
\begin{align*}
     \sqrt{\ln{\left(\frac{4 MT}{\hat{p}\Delta }\right)}}>\sqrt{\frac{1}{2}} \frac{\gamma(     \Delta)}{\sqrt{\Delta}}
\end{align*}
provided $\hat{p}=\min\left(1, \alpha \Delta^{\ln(1/\Delta)}\right)$  where $\alpha>0$. By substitute $M=1$ and simplifying we obtain
\begin{align*}
   2\Delta\left(\ln\left(\frac{4T}{\alpha }\right)+ \ln(1/\Delta)+\left(\ln(1/\Delta)\right)^{2}\right)>  \gamma^2( \Delta),
\end{align*}
for $\Delta<\Delta_0$. Hence the first claim is true as the left hand side of the above converges to $0$ as $\Delta\to 0$.
\end{proof}
\begin{Theorem} Let $\beta$ be the positive volatility parameter appearing in \eqref{EmpMod} and $\hat{\beta}_{\max}$ as in Definition \ref{defi}. As $\Delta\to 0$, $\hat{\beta}_{\max}$ converges to $\beta$ in probability.
\end{Theorem}
\begin{proof}
\noindent First note that, due to the finite activity jump, $SD^2 =O(\Delta)$ as $\Delta\to 0$. Thus by Theorem \ref{2.6} we know that the law of $\hat{\beta}_{\max}$ on $(0,\infty)$ is tight as $\Delta\to 0$ provided $\beta\neq 0$. That is, given a positive $\epsilon$, there is a pair $0<l_\epsilon <u_\epsilon$ such that
\begin{align}\label{tight}
P(\hat{\beta}_{\max} \notin [l_\epsilon, u_\epsilon])<\epsilon \textrm{ for all } \Delta>0.
\end{align}


\noindent Note that log return, i.e., the increment of log of asset price in $\Delta$ time span is
\begin{align} \label{lr}
\ln(S_{t})-\ln(S_{t-\Delta}) = \ln\left(1+ \frac{S_{t}-S_{t-\Delta}}{S_{t-\Delta}}\right) = \frac{S_{t}-S_{t-\Delta}}{S_{t-\Delta}} - O \left(\frac{S_{t}-S_{t-\Delta}}{S_{t-\Delta}}\right)^2.  
\end{align}
On the other hand, $\frac{S_{t}-S_{t-\Delta}}{S_{t-\Delta}}$ is the simple return on the same time span and goes to zero as $\Delta \to 0$ provided $t$ is not a jump time. Again, the discrete model \eqref{ret} implies that the square of simple return as well as the average  of simple return both are of order $O(\Delta)$ in probability. We also recall that $\Delta \ll \sqrt{\Delta} \ll \gamma(\Delta)$ (Lemma \ref{lemma2} (2)).
Let $x$ be a positive constant.
Therefore, by letting $\Delta\to 0$, almost surely the asymptotic jump identification by using  $x\gamma(\Delta)$ as the threshold on the distance of simple return from its average, will be identical to that by  using the same threshold on the absolute log return. Hence due to the asymptotic of Lemma \ref{lemma2}, Corollary 2 in \cite{man08}
can be applied with threshold $x \gamma(\Delta)$, and thus the corresponding estimator of $\beta$ converges to $\beta$ in probability. As all such estimators converge o the same limit, using \eqref{tight}, we may consider $ \hat{\beta}_{\max} \gamma(\Delta)$ as a threshold and conclude that the corresponding estimator of $\beta$ converges to $\beta$ in probability. However, the  estimator corresponding to the threshold $ \hat{\beta}_{\max} \gamma(\Delta)$ is   $\hat{\beta}_{\max}$ by Definitions \ref{defi1} and \ref{defi}. Hence the proof.
\end{proof}

\subsection{Finite sample properties} In this subsection, we study the finite sample performance of the estimator and threshold, presented in Definition \ref{defi} via some numerical experiments. The experiments involve  accuracy evaluation of jump detection
for a given family of simulated data. To be precise the accuracy measure is given by
\begin{align*}
\text{Accuracy}:=\frac{\text{True Positive + True Negative}}{\text{True Positive + False Positive + True Negative + False Negative}}. \end{align*}
For simulation, Merton's jump-diffusion (MJD) model \eqref{EmpMod} is considered with many different values of $\beta$ and by fixing all other parameters to its typical values. In particular we fix $\mu=0.1$, $\Lambda=100$ and $F$ is the cdf of $Lognormal(-\frac{\delta^2}{ 2}, \delta^2) -1$, with $\delta=0.0055$.
Furthermore, each simulated time series data is of length $18000$ and has granularity $\Delta=1/18000$. For each value of $\beta$ the average accuracy of $1000$ simulations is presented in a plot and a table below. In those, the performance of maximal threshold is also contrasted with few fixed thresholds. The fixed thresholds are chosen following the suggestions given in Section 5 of \cite{man08}.
\begin{minipage}{\linewidth}
	\centering
	\begin{minipage}{0.48\linewidth}
\begin{figure}[H]
\includegraphics[width=\linewidth]{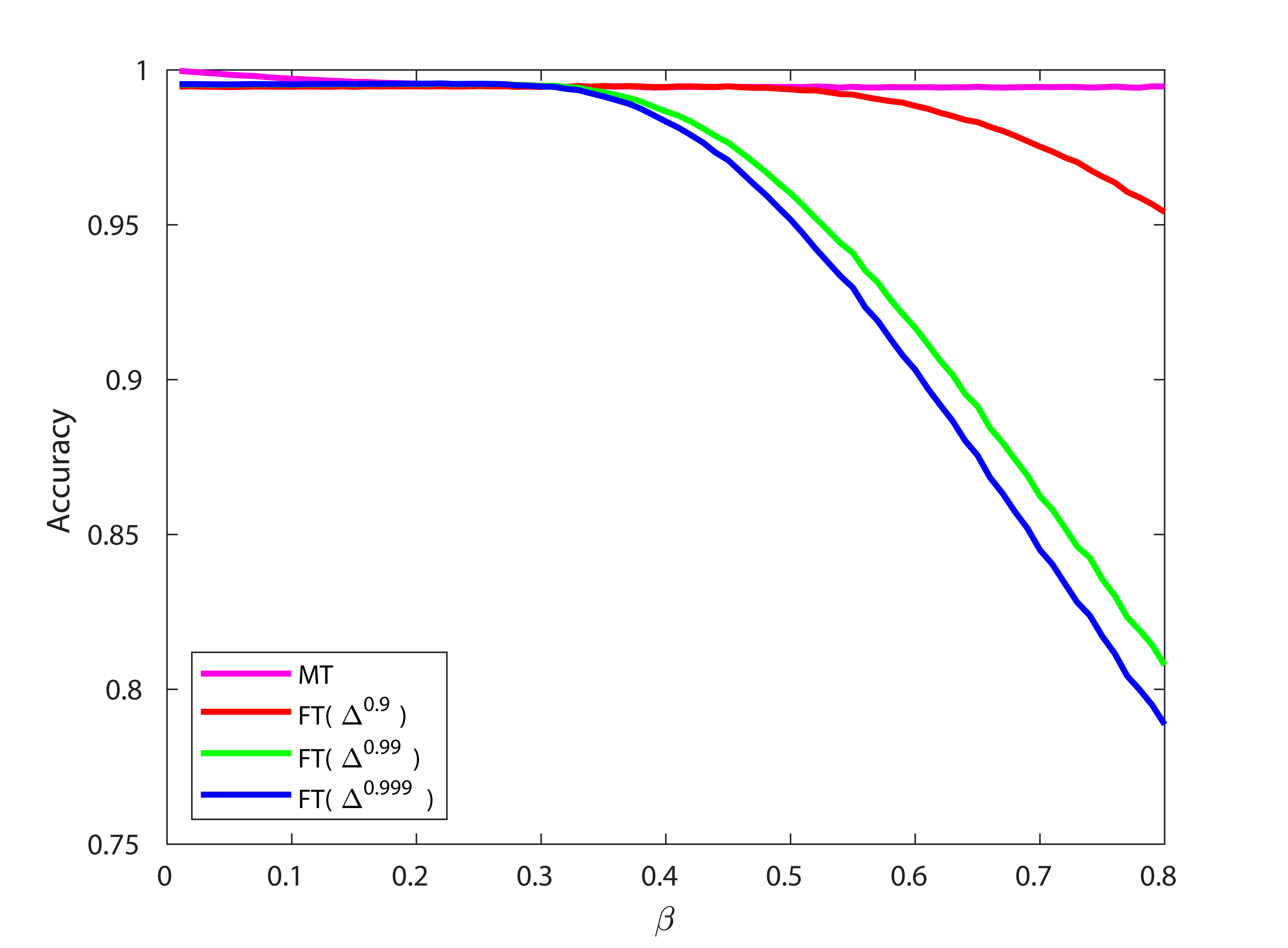}
\vspace*{-0.3 in}\\
\caption{Jump-detection Accuracy using Maximal threshold (MT) and Fixed threshold (FT) with values $\Delta^{0.9}$, $\Delta^{0.99}$ and $\Delta^{0.999}$.
}
\label{accmjd}
\end{figure}
\end{minipage}
\hspace{0.01\linewidth}
\begin{minipage}{0.48\linewidth}
\begin{table}[H]
	\centering
\begin{tabular}{|c|c|ccc|}
		\hline
		\multicolumn{1}{|c|}{$\beta$}& \multicolumn{1}{|c|}{\small Using Maximal}& \multicolumn{3}{|c|}{Using Fixed Threshold}\\
		\cline{3-5}
		
		{values} & {\small Threshold}   &  {$\Delta^{0.9}$} &  {$\Delta^{0.99}$}  & {$\Delta^{0.999}$}  \\\hline
		0.01 & 0.9997 &  0.9946 & 0.9953 &0.9954\\
		0.1 &   0.9972 & 0.9946 & 0.9953 & 0.9954\\
		0.2 & 0.9955 & 0.9947 & 0.9954 &0.9955\\
		0.3 & 0.9948 & 0.9947 &0.9950 & 0.9948\\
		0.4 & 0.9946 & 0.9947 & 0.9870 & 0.9838\\
		0.5 &  0.9945 & 0.9938 & 0.9603 &0.9517\\
		0.6 & 0.9945 &0.9885 &0.9164 &0.9027\\
		0.7 & 0.9945 & 0.9755 & 0.8632 &0.8459\\
		0.8 & 0.9944 & 0.9541 & 0.8083 & 0.7887\\
    \hline
\end{tabular}
\caption{Mean accuracy of jump-detection using 1000 simulations of MJD model for different $\beta$ and threshold values.}
\label{tab1}
\end{table}
	\end{minipage}
\end{minipage}
In Figure \ref{accmjd} the horizontal axis represents increasing values of the volatility parameter $\beta$ while the vertical axis represents the accuracy measure in jump detection by a threshold method. Each line plot shows how accuracy changes with $\beta$ corresponding to a threshold selection. Apart from the maximal threshold, three different fixed thresholds have been considered. Those are $\Delta^{0.9}$, $\Delta^{0.99}$ and $\Delta^{0.999}$. The maximal thresholds are obtained by fixing $\hat{p}=0.01$. Some of the numerical values are reported in Table \ref{tab1}.
It is evident from Figure \ref{accmjd} and Table \ref{tab1} that the accuracy measure diminishes in case of higher volatility if fixed thresholds are used for jump detection. However, the same  does not decline if maximal threshold is used.\\

\begin{figure}
\includegraphics[width=\linewidth]{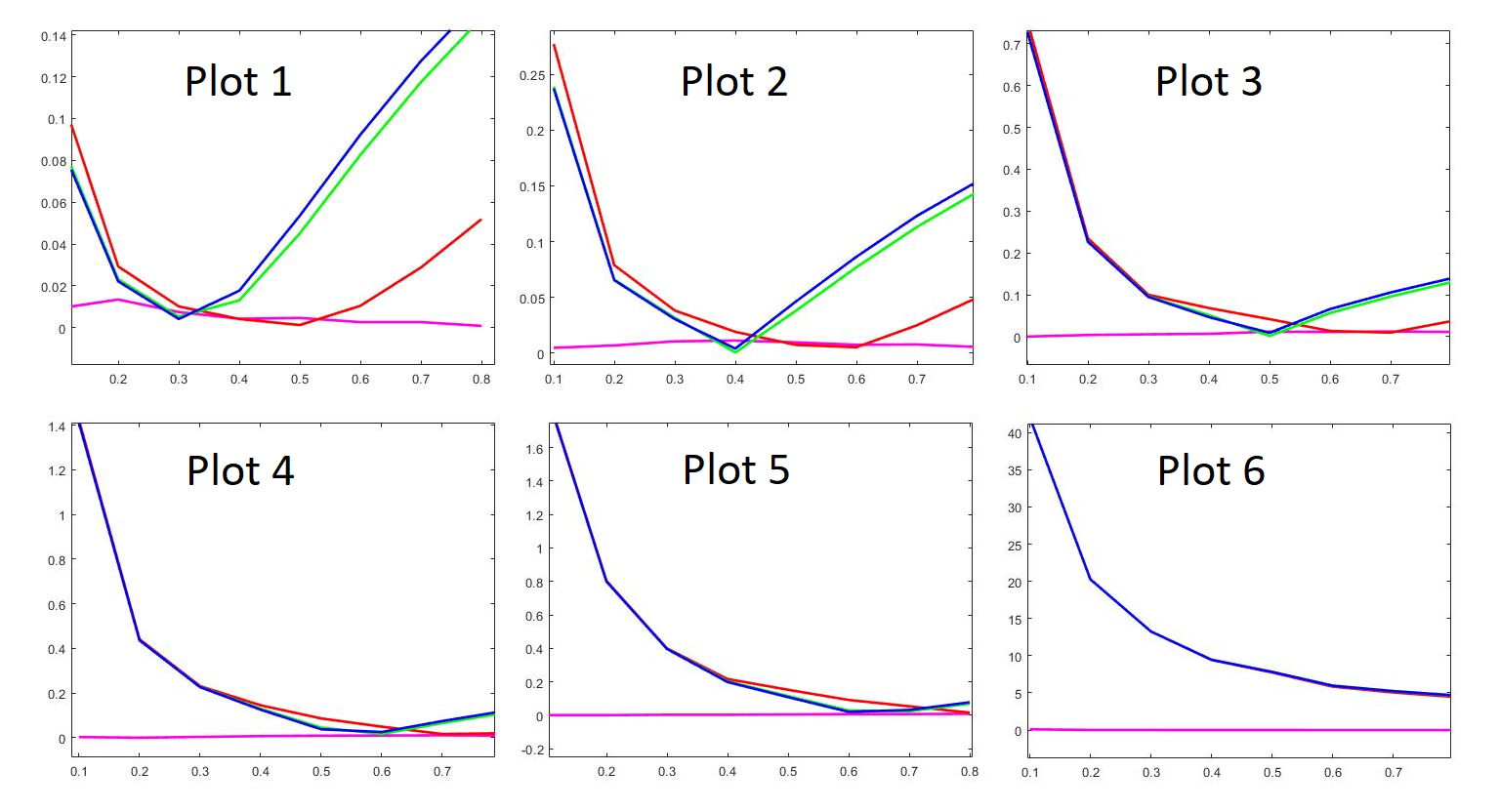}
\caption{Relative error vs true value in volatility estimation for $\delta=$ 0.0055, 0.01, 0.02, 0.03, 0.04, and 0.5 respectively. The color legend of all six plots are identical to that of Figure \ref{accmjd}.}
\label{estisix}
\end{figure}

\noindent In a further experiment we compute the relative errors of the volatility estimators obtained using above four thresholds. We study by varying $\beta$ in $[0,0.8]$ and taking six different $\delta$ values. Plots 1 to 6 in Figure \ref{estisix} correspond to $\delta$ values 0.0055, 0.01, 0.02, 0.03, 0.04, and 0.5 respectively. Clearly the range of error is larger for larger $\delta$ in the plots. This can be explained in the following manner. Larger $\delta$ produces more jumps having smaller size. This causes higher number of false negative in jump detection using fixed thresholds. On the contrary, since maximal threshold depends linearly on $\beta$, it can detect these small jumps if $\beta$ is small. On the other hand, the typical small jumps are still larger than the return size due to the diffusion noise provided $\beta$ is not too large. Thus, by the fixed threshold the estimation of small volatility is heavily affected by the presence of too many outliers coming from the small jumps. This explains the initial decline of relative error in small volatility estimation by fixed threshold when $\beta$ grows. In Plots 1 to 5 a further rise of error with $\beta$ values can be seen for higher $\beta$. This is attributed to the increase of false positive in jump detection by the fixed thresholds. On the contrary, since maximal threshold grows linearly with $\beta$, it does not misclassify large returns coming from diffusion noise as jumps.

\noindent The above mentioned experiments and explanation justifies the use of maximal threshold in jump detection in place of fixed threshold method. Of course there are many other approaches reported in the literature those can overcome the limitations of the fixed threshold approach. However, most of those are iterative methods and computationally expensive. In view of this, we apply maximal threshold for removing jumps from the time series and obtain the continuous part of the time series. Subsequently, we investigate the inference of the continuous part in the following section.

\section{Historical volatility: squeeze and expansion duration}\label{duration}
\noi In this and subsequent sections we write $\hat{\beta}$ and $\hat{c}$ for denoting $\hat{\beta}_{\max}$ and $\hat{c}_{\max}$ respectively. Thus $\hat{c}$ is the maximal threshold value for identifying the jump discontinuities. For each $i=1,\ldots, N$, we define
$$\hat r(i) :=
\begin{cases}
r(i) \textrm{ if } |r(i) -\bar{r}| < \hat{c}\\
\bar r \textrm{ else.}
\end{cases}
$$
Hence, $r(j)=\hat r(j) + (r(j)-\bar{r}) \mathds{1}_{ [ \hat{c},\infty)}(|r(j) -\bar{r}|)$. Thus, $\hat{r}=\{\hat r(i) \mid i=1,\ldots, N\}$ represents the simple return of the continuous part of the time series after removing the jump discontinuities. We use $\hat{r}$ to derive the historical volatility time series below.
\begin{Definition}[$ \hat{\mu},\hat{\sig}$]\label{mu:sighat}
For a fixed window size $n$, the moving average $\{m(k)\}_{k=n}^N$ and the sample standard deviation $\{\sig(k)\}_{k=n}^N$ are given by
	\begin{align}
	m(k) &:= \frac{1}{n}\sum_{i=0}^{n-1} \hat r(k-i),\label{movavg}\\
	\sig(k) &:= \sqrt{\dfrac{1}{n-1} \sum_{i=0}^{n-1} \left(\hat r(k-i)\right)^2 - \dfrac{n}{n-1} m(k)^2}\label{movstd},
	\end{align}
	for $k\geq n$. The empirical volatility $\hat{\sig} = \{\hat{\sig}(k)\}_{k= n}^N$ is given by $\hat{\sig}(k) := \dfrac{\sig(k)}{\sqrt{\Delta}}$. Similarly, the empirical drift $\hat{\mu}=\{\hat{\mu}(k)\}_{k= n}^N$ is given by $\hat{\mu}(k) := \dfrac{m(k)}{\Delta}$.
\end{Definition}
\noindent We also recall two more standard definitions from \cite{CB}.
\begin{Definition}\label{ecdf}
	Let $y=\{y(k)\}_{k=1}^m$ be a random sample of size $m$ from a real valued distribution. Then the empirical cumulative distribution function or ecdf $\hat{F}_y$ is defined as
	 \begin{equation*}
	\hat{F}_y(x):=\frac{1}{m}\ds\sum_{k=1}^{m}\mathds{1}_{[0,\infty)}(x-y(k)),
	\end{equation*}	
	where given a subset $A$, $\mathds{1}_{A}$ denotes the indicator function of $A$.
\end{Definition}
\begin{Definition}[$p$-percentile]\label{p:percentile}
	Let $\hat{F}_y$ be the ecdf of  $y=\{y(k)\}_{k=1}^m$. Then for any $p\in (0,1)$, the $100p$ percentile of $y$, denoted by $\hat{F}_y^{\leftarrow}(p)$, is defined as
	\begin{equation*}
	\hat{F}_y^{\leftarrow}(p):=\inf\bigg\{x\big|\hat{F}_y(x)\geq p\bigg\}.
	\end{equation*}
\end{Definition}
\noindent Following \cite{DG19}, with a particular $p$, the $100p$ percentile of $\hat{\sig}$ is used as the threshold for identifying the squeeze of the Bollinger band of return series. The precise definition is recalled below.
\begin{Definition}[$p$-squeeze] Given a $p\in (0,1)$, an asset is said to be in $p$-squeeze at $k$-th time step if the empirical volatility $\hat{\sig}(k)$, as defined above, is not more than $\hat{F}_{\hat{\sig}}^{\leftarrow}(p)$.
\end{Definition}
\noi We also recall from \cite{DG19}, the sojourn time duration of the $p$-squeeze  below.
\begin{Definition}\label{SqD}
	By following the convention of $\min\emptyset =+\infty$, for a fixed $p \in (0,1)$ and a given time series $\{\hat{\sig}\}_{k=n}^N$, let $\{(a_i,b_i)\}_{i=1}^\infty$ be an extended real valued double sequence given by \begin{equation*}
	\left\{ \begin{array}{lll}
	a_0=n\\
	b_{i-1}:=\min\{k\geq a_{i-1}|\hat{\sig}(k)>\hat{F}_{\hat{\sig}}^{\leftarrow}(p)\}\\
	a_i:=\min\{k\geq b_{i-1}|\hat{\sig}(k)\leq \hat{F}_{\hat{\sig}}^{\leftarrow}(p)\},
	\end{array}\right.
	\end{equation*}
for $i=1,2,\ldots$. Then the collection of sojourn time durations for the p-squeezes is $d(\hat{\sig}; p):=\{d_i\}_{i=1}^L$, where  $d_i:=b_i-a_i$ and $L:=\max\{i|b_i<\infty\}$, provided $L\geq 1$. In particular, we call $d_i$ as the $i$-th entry of $d(\hat{\sig}; p)$ and $L$ as the length of $d(\hat{\sig}; p)$.
\end{Definition}

\begin{figure}[h] \includegraphics[width=0.6\linewidth]{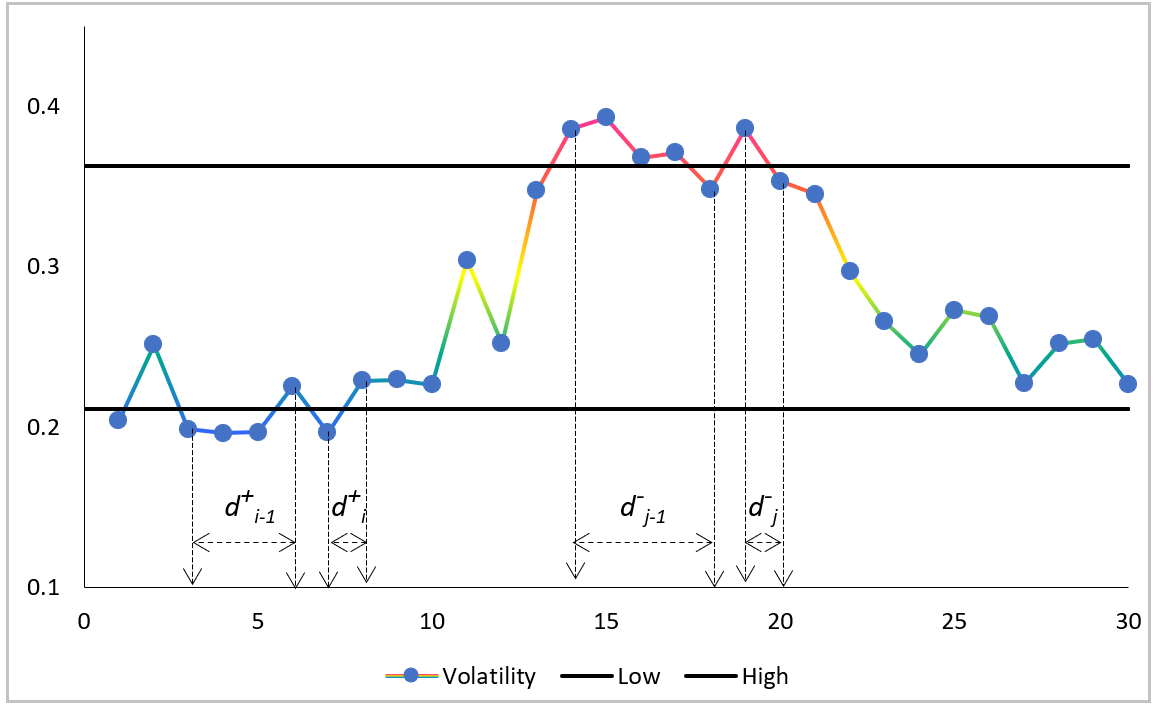} \caption{Illustration of duration of squeeze ($d^+$) and expansion ($d^-$)}\label{dur} \end{figure}

\noi We note that one must multiply each $d_i$ by $\Delta$ to obtain the squeeze duration in year unit. The following lemma helps to gain notational advantage in denoting the $p$-expansions, to be defined in Theorem \ref{lem2}.
\begin{Lemma}\label{lem1}
Given a time series  $y=\{y(k)\}_{k=1}^m$, and $p\in (0,1)$,
\begin{enumerate}
    \item[(i)] $-\hat{F}_{-y}^{\leftarrow}(p)=\hat{F}_y^{\leftarrow}(1-p+),$
    \item[(ii)] and if $p\in (0,1)\setminus \hat F_{y}(\mathbb{R})$, then $-\hat{F}_{-y}^{\leftarrow}(p)=\hat{F}_y^{\leftarrow}(1-p),$
\end{enumerate}
\end{Lemma}
 \begin{proof}
Firstly, note that $\hat{F}_{y}^{\leftarrow}(p)$ exists for any given time series $y$, and $p\in (0,1)$. Let $x=-\hat{F}_{-y}^{\leftarrow}(p)$, i.e., $-x=\hat{F}_{-y}^{\leftarrow}(p)$. For all $\vp>0$ the left and the right inequalities of the following are obtained by using the Definition \ref{p:percentile}; and non-decreasing right continuity of $\hat F$ respectively
\begin{align*}
   \hat{F}_{-y}(-x-\vp)< & p \leq \hat{F}_{-y}(-x).
\end{align*}
Again, by using Definition \ref{ecdf}, the above inequality can be rewritten as
\begin{align*}
  \frac{card\{ k \mid  -y(k)\leq -x -\vp\}}{m}< & p \leq \frac{card\{ k \mid  -y(k) \leq -x \}}{m},
  \end{align*}
  which can be equivalently written as
  \begin{align*}
    \frac{card\{ k \mid  y(k)\geq x +\vp\}}{m}< & p \leq \frac{card\{ k \mid  y(k)\geq x \}}{m}.
  \end{align*}
As length of the time series $y$ is $m$, the terms on both sides of the above inequalities can be rewritten using the complementarity as
  \begin{align*}
    1-\frac{card\{ k \mid  y(k)< x +\vp\}}{m}< & p \leq 1-\frac{card\{ k \mid  y(k)< x \}}{m}
  \end{align*}
for every $\vp>0$. A multiplication by $-1$ followed by an addition of $1$ to each of the terms give
  \begin{align*}
     \frac{card\{ k \mid  y(k)< x +\vp\}}{m}> & 1-p \geq \frac{card\{ k \mid  y(k)< x \}}{m},
  \end{align*}
for each $\vp>0$, which is reordered as
\begin{align*}
\frac{card\{ k \mid  y(k)\leq x-\vp \}}{m}\leq  \frac{card\{ k \mid  y(k)< x \}}{m}\le & 1-p < \frac{card\{ k \mid  y(k)< x +\vp\}}{m} \leq \frac{card\{ k \mid  y(k) \leq x +\vp\}}{m}.
  \end{align*}
Using Definition \ref{ecdf}, we get \begin{align*}
\hat{F}_{y}(x-\vp)\le & 1-p < \hat{F}_{y}(x+\vp),
\end{align*}
for each $\vp>0$ and hence, by right continuity and piece-wise constancy of $\hat{F}_{y}$ we deduce by taking $\vp\downarrow 0$
  \begin{align*}
\hat{F}_{y}(x-)\le & 1-p < \hat{F}_{y}(x).
\end{align*}
Hence, $x= \lim_{\vp\downarrow 0}\hat{F}_y^{\leftarrow}(1-p+\vp)$ and (i) follows.

\noi Secondly, if $p$ is not in the range of $\hat F_{y}$, $p$ is strictly less than $\hat{F}_{-y}(-x)$. Therefore, a derivation as above would produce $\hat{F}_{y}(x-)< 1-p < \hat{F}_{y}(x)$, in other words, $x= \hat{F}_y^{\leftarrow}(1-p)$.
\end{proof}
\begin{Theorem}\label{lem2}
$d(-\hat{\sig}; p)$ is the collection of sojourn time duration for $p$-expansions, i.e., the duration when $\hat{\sig}(k)\ge \hat{F}_{\hat{\sig}}^{\leftarrow}(1-p)$, provided $p$ is not in the range of $\hat F_{\hat{\sig}}$.
\end{Theorem}
\begin{proof}
The proof is a direct application of Lemma \ref{lem1}. To see this, note that if $d(-\hat{\sig}; p) =\{d_i\}_{i=1}^L$, then $d_i=b_i-a_i$, where $a_i=\min\{k\geq b_{i-1}|-\hat{\sig}(k)\leq \hat{F}_{-\hat{\sig}}^{\leftarrow}(p)\}$ that is same as $\min\{k\geq b_{i-1}|\hat{\sig}(k)\geq \hat{F}_{\hat{\sig}}^{\leftarrow}(1-p)\}$ and $b_{i-1}= \min\{k\geq a_{i-1}|-\hat{\sig}(k)>\hat{F}_{-\hat{\sig}}^{\leftarrow}(p)\}= \min\{k\geq a_{i-1}|\hat{\sig}(k)<\hat{F}_{\hat{\sig}}^{\leftarrow}(1-p)\}$ for $i\ge 1$ and $a_0=n$.
\end{proof}
\noi We write $d(\pm\hat{\sig}; p)$ to denote either $d(\hat{\sig}; p)$ or $d(-\hat{\sig}; p)$.  The Figure \ref{dur} illustrates the duration concept introduced in Definition \ref{SqD} and Theorem \ref{lem2}. It is important to note that in Lemma \ref{lem1}, the range of $\hat F_{y}$, i.e., $\hat F_{y}(\mathbb{R})$ is $\{i/m\mid i=0,1,\ldots, m\}$ since $y=\{y(k)\}_{k=1}^m$. Thus $(0,1)\setminus \hat F_{y}(\mathbb{R}) = \bigcup_{i=1}^m (\frac{i-1}{m}, \frac{i}{m})$.

\begin{Remark}\label{Rem}
\noi (i) From the construction of $d(\pm\hat{\sig}; p)$ it is evident that while  $d(\hat{\sig}; p)$ captures the duration of visiting low volatility, $d(-\hat{\sig}; p)$ captures that of visiting high volatility when $p$ is smaller than half. Thus when combined together, they can capture the three regime scenario, namely, low, medium and high volatility switching dynamics if $p$ is considerably smaller than half. Furthermore, the general class of three regimes models has the following two binary regime subclasses, namely (1) where the medium and high regimes are identical, in other words low volatility occurs with low probability (LVLP),  (2) where the medium and low regimes are identical, in other words high volatility occurs with low probability (HVLP). To study LVLP models, naturally $d(\hat{\sig}; p)$ is relevant and not $d(-\hat{\sig};p)$ whereas to study HVLP models, $d(-\hat{\sig}; p)$ is more appropriate when $p$ is small. Instead of studying the ternary regime switching models which involve too many parameters, we would only consider the above mentioned special classes of binary regimes in this paper for inference purpose. We would also test the hypothesis of single regime model as given in \eqref{EmpMod}. After removing the jump term from model \eqref{EmpMod}, one obtains a geometric Brownian motion which is also known as the Black-Scholes-Merton model. To test the model hypothesis \eqref{EmpMod}, we would use both $d(\hat{\sig}; p)$ and $d(-\hat{\sig};p)$ in Section \ref{empirical}.
\end{Remark}

\section{A Discriminating statistic}\label{section3}
\subsection{Construction of the discriminating statistic}
\noi It is well known that the asset price data of long past has little relevance in modelling the price dynamics in recent time. On the other hand, the use of diffusion term loses its justification for high frequency data. This puts an upper bar on the length of the time series under consideration for inference of switching in diffusion term. As a result, for a practically relevant time series data, the length of $d(\pm\hat{\sig}; p)$ is considerably small. Therefore, a non parametric estimation of the entries of $d(\pm\hat{\sig}; p)$ using empirical cdf is not practicable as that would have a high standard error. Hence, only a collection of few descriptive statistics such as mean($\bar{d}$), standard deviation($s$), skewness($\nu$), kurtosis($\kappa$) of $d(\hat{\sig}; p)$ or $d(-\hat{\sig}; p)$ should be considered as those can reliably be obtained. Although for a theoretical model, the corresponding $d(\pm\hat{\sig};p)$ is a random sequence with random length, the corresponding descriptive statistic constitutes a random vector of fixed length. The sampling distribution of this vector would be compared with the particular value $(\bar{d}, s, \nu, \kappa)$ of the time series data for testing the model hypothesis. In view of this, we construct a discriminating statistic $\mathbf{T}=(T_1,T_2,\ldots,T_r)$ using first $r$ number of descriptive statistics of $d(\pm\hat{\sig}; p)$. To be more specific we choose
\begin{align*}
T_1&:=\frac{1}{L}\ds\sum_{i=1}^{L}d_i, &
T_2&:=\sqrt{\frac{1}{L-1}\ds\sum_{i=1}^{L}(d_i-T_1)^2},\\
T_3&:=\dfrac{\frac{1}{L}\ds\sum_{i=1}^{L}(d_i-T_1)^3}{T^3_2},  &
T_4&:=\dfrac{\frac{1}{L}\ds\sum_{i=1}^{L}(d_i-T_1)^4}{T^4_2}
\end{align*}
etc. The test statistic is constituted with durations which are directly correlated to the sojourn times of regime transitions. Despite, it is not obvious that the statistic would indeed capture those unobserved switching successfully. The potential failure of capturing transitions is due to the presence of diffusion noise. The effect of this randomness can be reduced by considering a larger moving window size ($n$) for defining $\hat{\sig}$ in Definition \ref{mu:sighat}. However, a larger window size ignores the intermittent transitions more often, which also enhances the inaccuracy. We fix $n=20$ now onward in the definition, in view of the popular choice by practitioners for computing the empirical volatility. Next we describe the procedure, adopted in this paper, of obtaining the sampling distribution of $\mathbf{T}$ under binary regime switching model hypothesis.

\subsection{Rejection criteria based on the statistic}\label{tetspres}
In this subsection we present a description of numerical computation of sampling distribution of $\mathbf{T}$ statistic under each model of the composite null hypothesis using Monte-Carlo simulation method, which is popularly known as typical surrogate approach following \cite{JT} (see \cite{JD} for \emph{composite hypothesis}). The rejection criterion is given below.
\begin{itemize}
\item[(a)]  A non-empty subclass $\mathscr{A}$ of $\Theta$, the class of models obeying the null hypothesis is fixed.
\item[(b)] For each $\theta \in \mathscr{A}$, $B$ number of time series $\{X^1,X^2,\ldots,X^B\}$ are sampled from the corresponding model $\theta$ with the same time step size as in $S$.
\item[(c)]  Depending on the LVLP or HVLP hypothesis, $d(\hat{\sig}; p)$ or $d(-\hat{\sig}; p)$ is considered respectively for defining $\mathbf{T}$. Let $\mathbf{t}^*:=\mathbf{T}(S)$ be the value of $\mathbf{T}$ of the observed data $S$ and  $\mathbf{t}^*=(t^*_1,t^*_2,\ldots,t^*_r)$.
Let $\mathbf{t}^i=(t^i_1,t^i_2,\ldots,t^i_r):=\mathbf{T}(X^i)$ for each $i=1,\ldots, B$. Then $\mathbf{t}_\theta$ denotes $\{\mathbf{t}^1, \mathbf{t}^2,\ldots,\mathbf{t}^B\}$, the set of values of $\mathbf{T}$ for $\{X^1,X^2, \ldots,X^B\}$ corresponding to each $\theta \in \mathscr{A}$.
\item[(d)] In order to measure the proximity of $\mathbf{t}^*$ with respect to the set $\mathbf{t}_\theta$, we define $g_B:\RR \to[0,1/2]$ given by  $g_B(x):=\frac{\min (x,(B-x))}{B} \mathds{1}_{[0,B]}(x)$ and a proximity measure
$$\alpha^\theta_r:=\ds\min_{j\leq r}\,g_B\left(\ds\sum_{i=1}^{B}\mathds{1}_{[0,\infty)}(t^*_j-t_j^i)\right).$$
\item[(e)] The measure of proximity of the data to a class $\mathscr{A}$ of models is defined as
\[\alpha_r=\ds\max_{\theta\in \mathscr{A}}\,\alpha_r^\theta.\]
\item[(f)] The hypothesis that $S$ is a sample from a model in the class $\mathscr{A}$, can be rejected
provided $\alpha_r$ is smaller than a predetermined value.
\end{itemize}

\begin{Remark}
We would like to emphasize that the above mentioned rejection has an empirical confidence level $100(1-2\alpha_r)$ when $r=1$, i.e., the statistic is one dimensional. However, this expression differs vastly from confidence level when the dimension of statistic is large. This fact is commonly known as the ``curse of dimensionality". In other words for a given model $\theta$, the probability of observing the value of $\alpha^\theta_r$ to be smaller than a small value is not so small when $r$ is large. Since the curse is not so fatal for the dimension $r$ less than five, we consider a four dimensional statistic in this paper.
\end{Remark}

\section{Discretization of continuous time models}\label{Simulation}
\noi For testing of model hypothesis we consider a discrete time version of the continuous time theoretical asset price model given by
	  \begin{equation}
	        dS_t = \mu(X_{t-}) S_{t-}dt + \sig(X_{t-}) S_{t-}dW_t + S_{t-}dM_t \label{bin}
	    \end{equation}
for $t>0$ with $S_0>0$, where $\{X_t\}_{t\ge 0}$ is a two-state nonexplosive pure jump process. The time step of discrete version is taken identical to the granularity of the time series data. So far the binary regime-switching models are concerned, the switching in the continuous part could be either Markovian or semi Markovian. We consider both of these types in different subsections here. Prior to those, we consider the Merton's Jump Diffusion (MJD) model which has a single regime. Thus, we analyze three different composite model hypotheses, namely, (i)uni-regime MJD, (ii) Markov switching binary regime MJD, and (iii) semi-Markov switching binary regime MJD. The central idea of testing each such composite model hypothesis, as described in the previous section, is based on the simulation of the discrete version of continuous-time models selected from an appropriately chosen range of models satisfying the model assumption. In the next section we present an approximation procedure of identifying an appropriate model class of reduced dimension for a given data corresponding to each composite hypothesis.

\noi It is important to note that, since the test statistic $\mathbf{T}$ is computed after removing the jump discontinuities, it depends only on the continuous part of the data. Therefore for the inference purpose, it is sufficient to simulate only the continuous part of the models. To be more precise, instead of simulating \eqref{bin}, it is enough to simulate the following SDE for computing the sampling distribution of $\mathbf{T}$\begin{align}\label{binary}
dS_t=S_t\left(\mu(X_t)\,dt+\sig(X_t)\,dW_t\right),
\end{align}
where  $\{X_t\}_{t\geq 0}$  is a $\{1,2\}$-valued stochastic process and  $\mu(X_t),\sig(X_t)$ are the drift and the volatility coefficients. This observation helps to reduce computational complexity considerably. Let $\{0=t_0<t_1<\cdots<t_N\}$ be an equi-spaced partition of the time interval where $t_{i+1}-t_i=\Delta$ for $i=0,1,\ldots,N-1$ and $\Delta$ is the length of time step in year unit and same as the granularity of the empirical data. We use this convention throughout this paper.
\subsection{Uni-regime}
In this subsection we consider the model hypothesis \eqref{EmpMod} for some arbitrary model parameters $\mu, \beta, \Lambda$ and $F$.
After removing the jump term, the model reduces to
\begin{align}\label{bsm}
dS_t=S_t\left(\mu\,dt+\beta\,dW_t\right)~t> 0, ~ S_0>0.
\end{align}
Equation \eqref{bsm} has a strong solution of the form
\begin{align}\label{solbsm}
S_t=S_0\exp\left(\mu t-\frac{1}{2}\beta^2 t+\beta W_t\right),~t\geq 0.
\end{align}
The discretized  version of \eqref{solbsm} is given by
\begin{align}\label{dissolbsm}
 S_{t_0}=S_0, \tab S_{t_{i+1}}=S_{t_{i}}\exp\left((\mu -\frac{1}{2}\beta^2)\Delta +\beta\,Z_i\right) \textrm{ for }i=0,1,\ldots,N-1
\end{align}
where $\{Z_i\mid i=0, \ldots, N-1\}$ are independent and identically distributed (i.i.d.) normal random variables with mean $0$ and variance $\Delta$.

\subsection{Binary Markov regime}
After removing the jump term from \eqref{bin}, the model reduces to \eqref{binary} where $X$ denotes a Markov chain. Since, the continuous time Markov chain $X$ can be characterized by its instantaneous transition rate matrix $\lambda:=
\begin{pmatrix}
-\la_1 & \la_1\\
\la_2 & -\la_2
\end{pmatrix}$, the class of all possible models in \eqref{binary} can be identified with the following set $\Theta$ of all possible parameters
\begin{align}\label{class:mmgbm}
\Theta=\{\theta=(\mu(1),\sig(1),\la_1,\mu(2),\sig(2),\la_2)|\mu(i)\in\mathbb{R},\sig(i)>0,\la_i>0, i=1,2\}.
\end{align}
Using the expression of strong solution, the discrete version of models corresponding to each member of $\Theta$ is given by
\begin{align}\label{dissolmbsm}
\left. \begin{array}{ll}
S_{t_{i+1}} & =  S_{t_{i}}\exp\left(\left(\mu(X(i)) -\frac{1}{2}\sig^2(X(i))\right)\Delta +\sig(X(i))\,Z_i\right),\\
X(i+1) & = X(i)-(-1)^{X(i)}\,P_i,
\end{array}\right\}
\end{align}
where $\{P_i \mid i=1, \ldots, N-1\}$ are independent to $Z_j$ for all $j$ and for each given $X(i)$, the conditional distribution of $P_i$ is independent of $\{P_1, P_2, \ldots, P_{i-1}\}$ and follows $Bernoulli(\lambda_{X_i}\Delta )$, a Bernoulli random variable with $Prob( P_i=1\mid X(i))=\lambda_{X(i)}\Delta $, provided $\Delta \ll \min \{1/\lambda_{i} \mid i=1,2\}$. Here for each $i$, $Z_i$ is as in \eqref{dissolbsm}.
\subsection{Binary semi-Markov regime}
After removing the jump term from \eqref{bin}, the model reduces to \eqref{binary} which is dependent on a two state semi-Markov process $\{X_t\}_{t\geq 0}$. A semi-Markov process can be specified by its instantaneous transition rate function on $[0,\infty)$, given by \[
\lambda(y):=
\begin{pmatrix}
-\la_1(y) & \la_1(y)\\
\la_2(y) & -\la_2(y)
\end{pmatrix}~~\forall~ y\in [0,\infty).
\]
As before, the class of all possible models in \eqref{binary} can be identified with the following set $\Theta$ of all possible parameters
\begin{align*}
\Theta=\{\theta=(\mu(1),\sig(1),\la_1(\cdot),\mu(2),\sig(2),\la_2(\cdot))|\mu(i)\in\mathbb{R},\sig(i)>0,~\la_i(\cdot)>0,~ i=1,2\}.
\end{align*}
As before, the discrete version of the model corresponding to each member of $\Theta$ is given by
\begin{align}\label{dissolsbsm}
\left. \begin{array}{ll} S_{t_{i+1}} & =  S_{t_{i}}\exp\left(\left(\mu(X_i) -\frac{1}{2}\sig^2(X(i))\right)\Delta +\sig(X(i))\,Z_i\right),\\
\ X(i+1) & = X(i)-(-1)^{X(i)}\,P_i,\\
Y(i+1) & = \left(Y(i)+\Delta\right)\left(1-P_i\right),
\end{array}\right\}
\end{align}
where $\{Z_i\}_i$ are as in \eqref{dissolbsm} and $\{P_i \mid i=1, \ldots, N-1\}$ are independent to $Z_j$ for all $j$ and for each given pair $(X_i, Y_i)$, the conditional distribution of $P_i$ is independent of $\{P_1, P_2, \ldots, P_{i-1}\}$ and follows $Bernoulli(\lambda_{X(i)}(Y(i))\Delta )$, a Bernoulli random variable with $Prob( P_i=1\mid X(i), Y(i))=\lambda_{X(i)}(Y(i)) \Delta $, provided $\Delta \ll \min \{1/\lambda_{i}(y) \mid y\ge 0, i=1,2\}$. This discretization is obtained from the semi-martingale representation of the semi-Markov process, as in \cite{GhS}. The readers are referred to \cite{GhS} for more details about this representation of semi-Markov process.

\section{Empirical study}\label{empirical}
\noi We consider the time series data of eighteen different Indian stock indices with $5$-minute granularity during the time period starting from $1$-st December, 2016 and ending on $30$-th June, 2017. Assuming six hours of trading in each day and two hundred and fifty trading days in a year, we set $\Delta = \frac{5}{250\times 360} \approx 5.5\times10^{-5}$. In order to separate the jump discontinuities, we consider $\hat{p}=1\%$. Then we solve \eqref{para:est} numerically
as described in Remark \ref{rem1}.
\begin{table}[H]
    \centering
    \caption{Estimated parameters of 5-min (1/12/16 - 30/06/17) data of 18 Indian  stock indices}
    \label{para}
\begin{tabular}{|l|l|rrc|rr|}
\hline
\multicolumn{2}{|c|}{Index} & \multicolumn{1}{c}{ $\hat{\beta}$} & \multicolumn{1}{c}{	$\hat{\Lambda}$} & \multicolumn{1}{c|}{ $V$} &  \multicolumn{1}{c}{$\bar{\hat{ \mu}}$}&  \multicolumn{1}{c|}{	$\bar{\hat {\sigma}}$}\\ \cline{1-2}
Code &	Name &  (in \%)	& 	& (in $10^{-5}$) &  (in \%)	&  (in \%)	\\ \hline
I01 &  NIFTY 100          &   7.42   &   132.13   &   2   &   6.48   &  6.90  \\
I02 &  NIFTY 200          &   7.55   &   127.35   &   2   &   8.66   &  7.00  \\
I03 &  NIFTY 50           &   7.35   &   132.18   &   2   &   7.73   &  6.88  \\
I04 &  NIFTY 500          &   7.25   &   143.42   &   2   &  10.50   &  6.70  \\
I05 &  NIFTY BANK         &  10.40   &   127.28   &   6   &  25.39   &  9.75  \\
I06 &  NIFTY COMMODITY    &  10.24   &   114.45   &   3   &  -5.38   &  9.48  \\
I07 &  NIFTY ENERGY       &  10.94   &   148.23   &   4   & -13.16   & 10.22  \\
I08 &  NIFTY FIN. SER.    &   9.80   &   128.96   &   5   &  24.70   &  9.16  \\
I09 &  NIFTY FMCG         &  11.90   &   167.56   &   5   &  17.57   & 11.04  \\
I10 &  NIFTY INFRA        &  10.76   &   103.17   &   3   &  12.06   & 10.02  \\
I11 &  NIFTY IT           &  11.27   &   156.26   &   4   &   6.17   & 10.34  \\
I12 &  NIFTY MEDIA        &  14.65   &   119.28   &   6   &  12.96   & 13.65  \\
I13 &  NIFTY METAL        &  16.25   &    99.94   &   6   & -20.77   & 15.12  \\
I14 &  NIFTY MNC          &   9.37   &    77.34   &   3   &  20.30   &  8.67  \\
I15 &  NIFTY PHARMA       &  12.02   &   154.67   &   5   & -26.18   & 11.07  \\
I16 &  NIFTY PSE          &  10.27   &   146.70   &   3   &  -9.34   &  9.48  \\
I17 &  NIFTY REALTY       &  19.54   &   114.41   &  11   &  63.28   & 18.09  \\
I18 &  NIFTY SERVICE SEC. &   8.40   &   143.47   &   3   &  17.45   &  7.87  \\ \hline
\end{tabular}
\end{table}
\noi The numerical approximations of $\hat{\beta}$, $ \hat{\Lambda}$ and $V$ for each index data are given in Table \ref{para}. Each row of Table \ref{para} corresponds to an index, whose name is mentioned in the second column with its code in the first column. Using the $\hat \beta $ value, we obtain the $\hat c$ value for each index using \eqref{c}. Then using the value of $\hat c$ we derive the time series $\hat \mu$ and $\hat \sigma$ according to the Definition \ref{mu:sighat}. In last two columns, Table \ref{para} enlists the empirical long run average drift $\bar{\hat{\mu}}$ and the empirical long run average volatility $\bar{\hat{\sigma}}$ for each index.\\

\noindent Note that the hypothesis of our interest is composite in nature (see \cite{JD} for \emph{composite hypothesis}). In other words, we do not fix any parameter value in the null hypothesis. This results in consideration of models with parameters coming from a high-dimensional linear space. For the sake of reduction of dimension, it is necessary to add some other natural criteria on parameters. In principle, those criteria should  put direct and easily calculable constraints on the parameter set of the class of models. Following the approach of \cite{DG19}, some constraints are fixed and presented in the following two definitions.
\begin{Definition}[$\mathscr{C}$-class] \label{padcls} Given a time series data, a regime switching model is said to be in $\mathscr{C}$-class of models if the model satisfies the following properties.
	\begin{itemize}
		\item[i.] The long run average of drift coefficient of the continuous part matches with the time average of empirical drift $\hat{\mu}$ of the data.
		\item[ii.] The long run average of volatility process for the model matches with the time average of empirical volatility $\hat\sig$ of the data.
	\end{itemize}
\end{Definition}
\noi In addition to this, we introduce two other subclasses, $\mathscr{C}^+_p$ and $\mathscr{C}^-_p$ of $\mathscr{C}$ to include LVLP and HVLP models respectively.
\begin{Definition}[$\mathscr{C}^{\pm}_p$-class]\label{padcls+}
	Given a time series data and a fixed $p\in (0,1)$, a regime switching model in $\mathscr{C}$ is said to be in
\begin{itemize}
\item	$\mathscr{C}^+_p$-class of models if the long run proportion of time that the volatility process stays below $\hat{F}_{\hat{\sig}}^{\leftarrow}(p)$ is $p$,
\item $\mathscr{C}^-_p$-class of models if the long run proportion of time that the volatility process stays above $\hat{F}_{\hat{\sig}}^{\leftarrow}(1-p)$ is $p$,
	\end{itemize}	
provided the volatility process is not constant. We write $\mathscr{C}^\pm_p$ to denote either $\mathscr{C}^+_p$ or $\mathscr{C}^-_p$.
\end{Definition}
For computational purpose, given a time series $S$, firstly the $\mathscr{C}$- and $\mathscr{C}^\pm_p$-class of models satisfying each composite hypothesis are identified. Next a non-empty subclass $\mathscr{A}$ of $\mathscr{C}^\pm_p$ obeying the null hypothesis is fixed. We fix $p=15\%$ in the definition of $\mathscr{C}^\pm_p$, and $d(\pm\hat{\sig}; p)$ and thus the statistic $\mathbf{T}$ is evaluated with $p=15\%$ throughout this section. We have computed the $\mathbf{t}^*$ values using $d(\hat{\sig}; p)$ and $d(-\hat{\sig}; p)$ separately. The components of $\mathbf{t}^*$ for every index data are given in the columns of the Table \ref{Table1}.
\begin{table}[H]
	\centering
	\caption{$\mathbf{t}^*$ of the empirical data} \label{Table1}	
	\begin{tabular}{|l|rrrrr|rrrrr|}
	\hline
		\multicolumn{1}{|c|}{}& \multicolumn{5}{|c|}{Squeeze duration  $d(\hat{\sig}; p)$}& \multicolumn{5}{|c|}{Expansion duration $d(-\hat{\sig}; p)$}\\ \cline{2-11}
		Index & $L$ & {$t_1^{\ast}$} & {$t_2^{\ast}$} & {$t_3^{\ast}$} & {$t_4^{\ast}$} & $L$ & {$t_1^{\ast}$} & {$t_2^{\ast}$} & {$t_3^{\ast}$} & {$t_4^{\ast}$} \\
		\hline
 I01 &        157  & 10.66 &  11.31 &  1.17 &  3.41 &      142   & 11.78 &  10.66 & 1.07   & 3.85   \\
 I02 &        169  &  9.89 &  11.12 &  1.36 &  3.97 &      141   & 11.85 &  10.48 & 0.94   & 3.36   \\
 I03 &        158  & 10.58 &  10.89 &  1.10 &  3.29 &      149   & 11.22 &  10.50 & 1.09   & 3.85   \\
 I04 &        158  & 10.59 &  11.23 &  1.23 &  3.66 &      143   & 11.69 &  10.29 & 0.92   & 3.43   \\
 I05 &        158  & 10.59 &  11.69 &  1.38 &  4.03 &      135   & 12.39 &   9.80 & 0.70   & 3.42   \\
 I06 &        168  &  9.95 &  10.55 &  1.49 &  4.62 &      136   & 12.29 &  10.12 & 0.87   & 3.65   \\
 I07 &        165  & 10.14 &  11.29 &  1.58 &  4.80 &      141   & 11.87 &   9.21 & 0.60   & 3.29   \\
 I08 &        172  &  9.72 &  10.85 &  1.56 &  4.66 &      121   & 13.82 &  10.39 & 0.77   & 3.74   \\
 I09 &        179  &  9.35 &  10.18 &  1.58 &  5.00 &      136   & 12.29 &  10.27 & 0.76   & 3.31   \\
 I10 &        176  &  9.50 &  11.69 &  1.75 &  5.49 &      128   & 13.06 &  11.72 & 1.28   & 5.17   \\
 I11 &        159  & 10.52 &  11.37 &  1.19 &  3.35 &      127   & 13.17 &  10.04 & 1.32   & 6.97   \\
 I12 &        174  &  9.61 &   9.50 &  1.22 &  3.91 &      122   & 13.70 &  10.87 & 0.60   & 2.80   \\
 I13 &        187  &  8.94 &  10.58 &  1.91 &  6.49 &      121   & 13.69 &   9.50 & 0.36   & 2.39   \\
 I14 &        178  &  9.40 &  10.64 &  1.54 &  4.67 &      128   & 13.02 &  10.01 & 0.68   & 3.31   \\
 I15 &        174  &  9.61 &  11.21 &  1.56 &  4.56 &      125   & 13.38 &  11.91 & 1.69   & 8.42   \\
 I16 &        148  & 11.30 &  12.66 &  1.30 &  3.80 &      140   & 11.94 &  10.38 & 0.86   & 3.13   \\
 I17 &        183  &  9.14 &  10.44 &  1.84 &  6.03 &      110   & 15.21 &  11.10 & 0.81   & 3.87   \\
 I18 &        171  &  9.78 &  11.10 &  1.34 &  3.82 &      119   & 14.05 &   9.72 & 0.50   & 3.42   \\
\hline
\end{tabular}
\end{table}
With the choice of $p=15\%$, the binary regime model classes $\mathscr{C}^+_p$ and $\mathscr{C}^-_p$ include the LVLP and the HVLP models respectively.
For testing LVLP model hypothesis, $d(\hat{\sig}; p)$ is considered to define $\mathbf{T}$; else for HVLP models, $d(-\hat{\sig}; p)$ is considered for defining $\mathbf{T}$. The $\alpha$ values are obtained for each cases, namely LVLP and HVLP respectively.
\subsection{Uni-regime model}
In this subsection we consider the model hypothesis \eqref{EmpMod} of uni-regime MJD process. For each index in Table \ref{para} and \ref{Table1}, we set our null hypothesis,
\begin{align*}
H_0:\text{the time series is in $\mathscr{C}$-class of  \eqref{EmpMod}.}
\end{align*}
It is easy to see that, there is a unique choice of $\mu$ and $\beta$ so that \eqref{EmpMod} is in the $\mathscr{C}$-class, as $\mu$ and $\beta$ are, by using Definition \ref{padcls} (i)-(ii), $\mu=\bar{\hat{\mu}}$ and $\beta=\bar{\hat{\sig}}$, where the bar sign represents the time average.\\

\noindent The following two box plots illustrate results from all 18 indices. While Figure \ref{gbmS} illustrates the sampling distribution of $T_1$ of $d(\hat{\sig}; p)$,  Figure \ref{gbmE} plots that of $d(-\hat{\sig}; p)$. Each box plot is obtained by simulating \eqref{dissolbsm} $200$ times. The dot plots represent $t^*_1$ obtained from the Table \ref{Table1}. As the dots appear non-overlapping with the box plots, the null hypothesis is rejected with $100\%$ confidence.

\begin{minipage}{\linewidth}
	\centering
	\begin{minipage}{0.48\linewidth}
		\begin{figure}[H]
			\includegraphics[width=\linewidth]{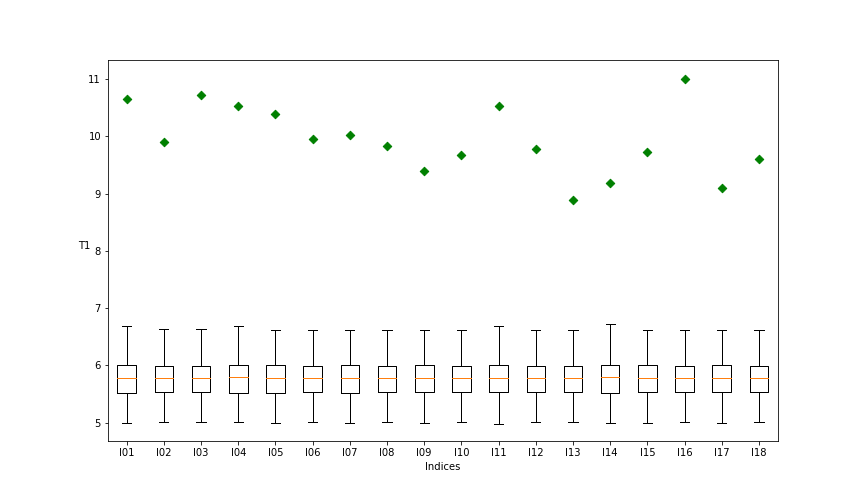}
			\caption{Sampling distribution of $T_1$ of $d(\hat{\sig}; p)$ under GBM hypothesis}\label{gbmS}
		\end{figure}
	\end{minipage}
	\hspace{0.02\linewidth}
	\begin{minipage}{0.48\linewidth}
		\begin{figure}[H]
			\includegraphics[width=\linewidth]{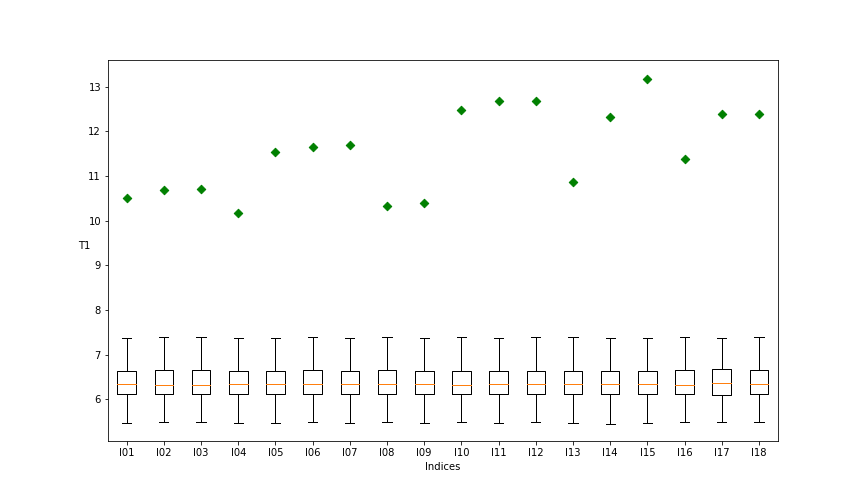}
			\caption{ Sampling distribution of $T_1$ of $d(-\hat{\sig}; p)$ under GBM hypothesis}\label{gbmE}
		\end{figure}
	\end{minipage}
\end{minipage}

\subsection{Binary Markov regime model}\label{5.2}
The parameter set of continuous parts of sub-classes $\mathscr{C}^+_p$ and $\mathscr{C}^-_p$ are subsets of $\Theta$ and would be derived, in this subsection, as the solution space of a system of equations. As the sojourn time distribution of state $i$ is Exp$(\la_i)$ for $i=1,2$ using Definition \ref{padcls+}, we have
\begin{equation}\label{lap1} \dfrac{\frac{1}{\la_1}}{\frac{1}{\la_1}+\frac{1}{\la_2}}=p \tab \textrm{or, } \la_1 = \left(\frac{1}{p}-1\right)\la_2. \end{equation}
Using Definition \ref{padcls}(i) the drift coefficients $\mu(i)$ satisfy the following relation
\begin{equation}\label{rel:mu} p\,\mu(1)+(1-p)\,\mu(2)=\bar{\hat{\mu}}. \end{equation}
Also using Definition \ref{padcls}(ii), the volatility coefficients $\sig(i)$ have the relation below
\begin{equation}\label{rel:si} p\,\sig(1)+(1-p)\,\sig(2)=\bar{\hat{\sig}}. \end{equation}
Thus the parameter set of continuous part of $\mathscr{C}^+_p$ is given by
\begin{align}\label{A+} \mathscr{A}^+:=\{\theta\in \Theta \mid \eqref{lap1}, \eqref{rel:mu}, \textrm{ and  } \eqref{rel:si} \textrm{ hold, and } \sig(1)\in [0, \hat{F}_{\hat{\sig}}^{\leftarrow}(p)]\}. \end{align}
Similarly, the parameter set of continuous part of $\mathscr{C}^-_p$ is given by \begin{align}\label{A-} \mathscr{A}^-:=\{\theta\in \Theta \mid \eqref{lap1}, \eqref{rel:mu}, \textrm{ and  } \eqref{rel:si} \textrm{ hold, and } \sig(1)\in [\hat{F}_{\hat{\sig}}^{\leftarrow}(1-p),\infty)\}.
\end{align}
For the reduction of computational complexity, we choose smaller sets $\mathscr{A}^+$ and $\mathscr{A}^-$ than in \eqref{A+} and \eqref{A-} respectively by fixing $\mu(1)=\mu(2)$. Thus now $\mathscr{A}^\pm$ is a subset of the solution space of four equations in six unknowns, or in other words, $\mathscr{A}^\pm$ can be viewed as a two-parameter family of models. The parameter $\lambda_1$ is varied by taking  $\frac{1}{\lambda_1\Delta} = 5,6, \ldots, 16$. That is $\lambda_1$ ranges from 1.1E+03 to 3.6E+03. On the other hand the range of parameter $\sigma(1)$ is not identical for $\mathscr{A}^\pm$ classes. However, we discretize those ranges with variable step size of one percentile. Or in other words, for $\mathscr{A}^+$, $\sigma(1)$ is chosen from the set $\{\hat{F}_{\hat{\sig}}^{\leftarrow}(i/100)\mid i=1,2,\ldots, \lfloor 100p\rfloor\}$ and for $\mathscr{A}^-$, $\sigma(1)$ is chosen from the set $\{\hat{F}_{\hat{\sig}}^{\leftarrow}(i/100)\mid i=\lceil 100(1-p)\rceil, \ldots, 100\}$.

\begin{figure}[H]
\includegraphics[width=0.6\linewidth]{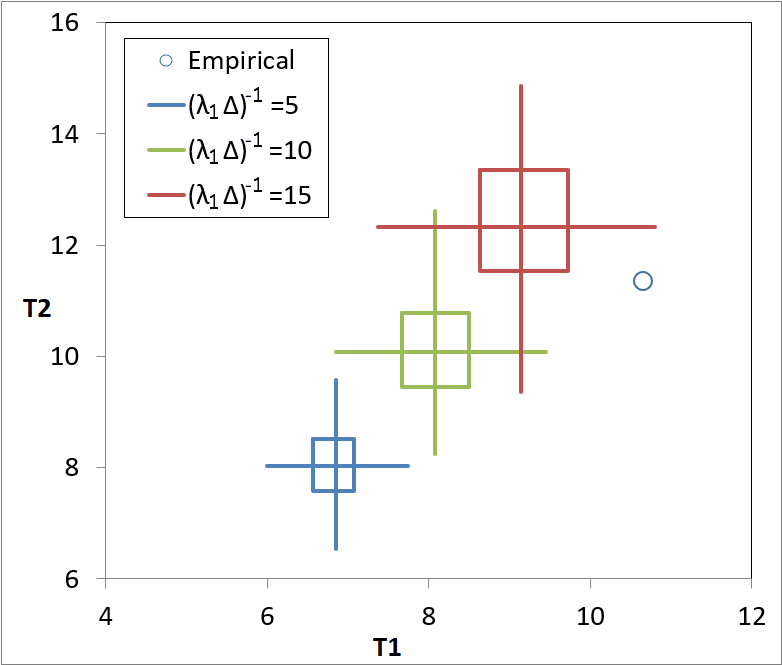}
\caption{Distribution of $(T_1,T_2)$ under LVLP Markov modulated MJD  models}\label{mgbm:t1t2}
\end{figure}
\noindent We illustrate the variability of sampling distribution of $(T_1,T_2)$ w.r.t. $\theta \in \mathscr{A}^+$ by considering the time series I01 in Figure \ref{mgbm:t1t2}. The circle plot represents $(t^*_1,t^*_2)$, the $(T_1,T_2)$ value of I01. There are three two-dimensional box plots corresponding to $\frac{1}{\lambda_1\Delta}$ equal to $5,10$ and $15$ respectively. In these $\sigma(1)$ value is set as $\hat{F}_{\hat{\sig}}^{\leftarrow}(p)$. These are merely 2-D projections of the 4-D distribution of $\mathbf{T}$. Nevertheless, the plots in Figure \ref{mgbm:t1t2} validate the sensitivity of the statistic w.r.t. the free parameters in $\mathscr{A}^\pm$. We recall that the above mentioned sensitivity is a necessary feature for successful calibration using a bootstrap approach. Now for each index in Table \ref{para} and \ref{Table1}, we set the null hypothesis,
\begin{align*}
H_0:\text{the law of time series is Markov modulated MJD \eqref{bin} with parameters of continuous part in $ \mathscr{A}^\pm$}.
\end{align*}
For each index, we compute the value of $\alpha_r$ as in the Subsection \ref{tetspres} for $r=1,2,3,4$ by simulating \eqref{dissolmbsm}. The results are presented in Table \ref{Table2}. The values show that for all indices, except I11 and I15, the proximity of the data to either LVLP or HVLP models, are more than 0.2. Furthermore, for most of the indices the HVLP models fit better than LVLP. Nevertheless, after removing the jumps, the proximity of most of the data to LVLP has increased (refer to Table 2 in \cite{DG19} for a comparison).
\begin{table}[ht]
	\centering
	\caption{The proximity of the index data to the classes of binary Markov regime models} \label{Table2}	
	\begin{tabular}{|l|rrrr|rrrr|}
		\hline
		\multicolumn{1}{|}{}& \multicolumn{4}{|c|}{For $\mathscr{A}^+$ using  $d(\hat{\sig}; p)$}& \multicolumn{4}{|c|}{For $\mathscr{A}^-$ using $d(-\hat{\sig}; p)$}\\ \cline{2-9}
		
		{Index} &   {$\alpha_1$} &  {$\alpha_2$} &   {$\alpha_3$} &   {$\alpha_4$} &   {$\alpha_1$} &  {$\alpha_2$} &   {$\alpha_3$} &   {$\alpha_4$}\\
		\hline
  I01 &   0.50 &   0.36 &   0.04 &   0.03 &     0.50 &    0.47 &  0.40    & 0.40 \\
  I02 &   0.50 &   0.45 &   0.12 &   0.06 &     0.48 &    0.38 &  0.38    & 0.38 \\
  I03 &   0.50 &   0.31 &   0.02 &   0.02 &     0.50 &    0.45 &  0.41    & 0.41 \\
  I04 &   0.50 &   0.35 &   0.06 &   0.05 &     0.50 &    0.46 &  0.46    & 0.46 \\
  I05 &   0.50 &   0.48 &   0.17 &   0.09 &     0.49 &    0.34 &  0.34    & 0.34 \\
  I06 &   0.50 &   0.33 &   0.22 &   0.16 &     0.49 &    0.47 &  0.35    & 0.31 \\
  I07 &   0.50 &   0.42 &   0.31 &   0.21 &     0.50 &    0.15 &  0.14    & 0.14 \\
  I08 &   0.50 &   0.47 &   0.24 &   0.15 &     0.49 &    0.38 &  0.36    & 0.36 \\
  I09 &   0.50 &   0.38 &   0.30 &   0.26 &     0.49 &    0.42 &  0.42    & 0.42 \\
  I10 &   0.50 &   0.48 &   0.36 &   0.26 &     0.50 &    0.44 &  0.35    & 0.30 \\
  I11 &   0.50 &   0.41 &   0.06 &   0.03 &     0.50 &    0.45 &  0.12    & 0.07 \\
  I12 &   0.50 &   0.22 &   0.03 &   0.03 &     0.50 &    0.41 &  0.33    & 0.26 \\
  I13 &   0.50 &   0.43 &   0.43 &   0.39 &     0.50 &    0.16 &  0.14    & 0.14 \\
  I14 &   0.50 &   0.45 &   0.26 &   0.16 &     0.49 &    0.45 &  0.45    & 0.42 \\
  I15 &   0.50 &   0.47 &   0.22 &   0.12 &     0.49 &    0.43 &  0.20    & 0.13 \\
  I16 &   0.50 &   0.43 &   0.12 &   0.07 &     0.50 &    0.49 &  0.44    & 0.35 \\
  I17 &   0.50 &   0.47 &   0.41 &   0.33 &     0.49 &    0.43 &  0.42    & 0.39 \\
  I18 &   0.50 &   0.45 &   0.10 &   0.05 &     0.50 &    0.27 &  0.27    & 0.27 \\
\hline
\end{tabular}
\end{table}
\subsection{Binary semi-Markov regime model}
First we specify the parameter set of continuous parts of sub-classes $\mathscr{C}^+_p$ and $\mathscr{C}^-_p$. The conditional cdf of holding time  distribution, given the state $i$, is $[0,\infty)\ni y\mapsto 1-e^{-\int_0^y \la_i(u) du}$. Hence the expected sojourn time at state $i$ is $E_i:=\int_0^\infty e^{-\int_0^y \la_i(u) du} dy$. Therefore, using Definition \ref{padcls+}, we have
\begin{equation}\label{lap2} \dfrac{E_1}{E_1+E_2}=p \tab \textrm{or, } E_2 = \left(\frac{1}{p}-1\right)E_1. \end{equation}
In addition to the above equation, the Definition \ref{padcls} implies that the parameters $\mu(i)$ and $\sig(i)$ should satisfy the equations \eqref{rel:mu} and \eqref{rel:si}. Thus the parameter set of continuous part of $\mathscr{C}^+_p$ is given by
\begin{align}\label{As+} \mathscr{A}^+:=\{\theta\in \Theta \mid \eqref{lap2}, \eqref{rel:mu}, \textrm{ and  } \eqref{rel:si} \textrm{ hold, and } \sig(1)\in [0, \hat{F}_{\hat{\sig}}^{\leftarrow}(p)]\}. \end{align}
Similarly, the parameter set of continuous part of $\mathscr{C}^-_p$ is given by
\begin{align}\label{As-} \mathscr{A}^-:=\{\theta\in \Theta \mid \eqref{lap2}, \eqref{rel:mu}, \textrm{ and  } \eqref{rel:si} \textrm{ hold, and } \sig(1)\in [\hat{F}_{\hat{\sig}}^{\leftarrow}(1-p),\infty)\}.
\end{align}
Note that $\mathscr{A}^\pm$ is not finite dimensional due to the presence of the functional parameters $\la_i(\cdot)$. Now for illustration purpose, $\mathscr{A}$ is chosen in the following manner. The holding time distribution of the state $i$ is assumed to follow  $\Gamma(k_i,\la_i)$ for $i=1,2$, where $\Gamma(k_i,\la_i)$ denote the gamma distribution with shape $k_i$ and rate $\la_i$. Then it follows from \cite{GhS} that $\la_i(y)$ is the hazard rate of $\Gamma(k_i,\la_i)$ and is given by
$\la_i(y) = \frac{\la_i^{k_i}y^{k_i-1}e^{-\la_iy}}{\Gamma(k_i)-\gamma(k_i,\la_iy)}$, where $\gamma$ is the lower incomplete gamma function. Since the expectation of $\Gamma(k_i,\la_i)$ is $\frac{k_i}{\la_i}$, it follows from \eqref{lap2}, that
\[\frac{\frac{k_1}{\la_1}}{\frac{k_1}{\la_1}+\frac{k_2}{\la_2}}=p, \tab \textrm{or, } \frac{k_2}{\la_2}=\left(\frac{1}{p}-1\right)\frac{k_1}{\la_1}.\]
In addition to these, as before, we further assume that $\mu(1)=\mu(2)$, and $k_1=k_2=k$ (say). Thus $\mathscr{A}^\pm$ is the solution space of five equations in eight unknowns or in other words $\mathscr{A}^\pm$ is a three parameter subfamily of $\Theta$. We vary $\lambda_1$ and $\sigma(1)$ as we do in the Subsection \ref{5.2}. The identified parameter $k$ is chosen from the set $\{2, \ldots, 15\}$ for excluding the Markov special case corresponding to $k=1$. For each index in Table \ref{para} and \ref{Table1}, we set the null hypothesis,
\begin{align*}
H_0:\text{the law of time series is semi-Markov modulated MJD \eqref{bin} with parameters of continuous part in $\mathscr{A}^\pm$}.
\end{align*}
For each index, we compute the value of $\alpha_r$ as in the Subsection \ref{tetspres} for $r=1,2,3,4$ by simulating \eqref{dissolsbsm}. The results are presented in Table \ref{Table3}.
\begin{table}[ht]
	\centering
	\caption{The $\alpha$-values for all the indices under binary semi-Markov regime model hypotheses} \label{Table3}
	\begin{tabular}{|l|rrrr|rrrr|}
		\hline
\multicolumn{1}{|}{}& \multicolumn{4}{|c|}{For $\mathscr{A}^+$ using  $d(\hat{\sig}; p)$}& \multicolumn{4}{|c|}{For $\mathscr{A}^-$ using $d(-\hat{\sig}; p)$}\\ \cline{2-9}
		{Index} &   {$\alpha_1$} &  {$\alpha_2$} &   {$\alpha_3$} &   {$\alpha_4$} &   {$\alpha_1$} &  {$\alpha_2$} &   {$\alpha_3$} &   {$\alpha_4$}\\
		\hline
  I01 &    0.50 &  0.49  &   0.29 &   0.20 &       0.50 &     0.40 &     0.36 &   0.33   \\
  I02 &    0.50 &  0.44  &   0.37 &   0.27 &       0.50 &     0.47 &     0.36 &   0.36   \\
  I03 &    0.50 &  0.41  &   0.22 &   0.21 &       0.50 &     0.41 &     0.41 &   0.41   \\
  I04 &    0.50 &  0.47  &   0.33 &   0.25 &       0.50 &     0.45 &     0.43 &   0.43   \\
  I05 &    0.50 &  0.45  &   0.38 &   0.31 &       0.50 &     0.36 &     0.36 &   0.30   \\
  I06 &    0.50 &  0.42  &   0.42 &   0.38 &       0.50 &     0.42 &     0.34 &   0.34   \\
  I07 &    0.50 &  0.44  &   0.44 &   0.41 &       0.50 &     0.14 &     0.14 &   0.14   \\
  I08 &    0.50 &  0.44  &   0.43 &   0.39 &       0.50 &     0.35 &     0.32 &   0.27   \\
  I09 &    0.50 &  0.43  &   0.41 &   0.39 &       0.50 &     0.39 &     0.39 &   0.39   \\
  I10 &    0.50 &  0.48  &   0.42 &   0.42 &       0.50 &     0.45 &     0.29 &   0.25   \\
  I11 &    0.50 &  0.47  &   0.28 &   0.16 &       0.50 &     0.44 &     0.12 &   0.05   \\
  I12 &    0.50 &  0.37  &   0.21 &   0.21 &       0.50 &     0.40 &     0.40 &   0.36   \\
  I13 &    0.50 &  0.49  &   0.46 &   0.43 &       0.50 &     0.22 &     0.22 &   0.22   \\
  I14 &    0.50 &  0.47  &   0.42 &   0.34 &       0.50 &     0.43 &     0.34 &   0.30   \\
  I15 &    0.50 &  0.50  &   0.36 &   0.31 &       0.50 &     0.38 &     0.18 &   0.10   \\
  I16 &    0.50 &  0.45  &   0.36 &   0.34 &       0.50 &     0.46 &     0.43 &   0.43   \\
  I17 &    0.50 &  0.46  &   0.46 &   0.38 &       0.50 &     0.34 &     0.34 &   0.33   \\
  I18 &    0.50 &  0.44  &   0.34 &   0.22 &       0.50 &     0.32 &     0.32 &   0.29   \\
\hline
\end{tabular}
\end{table}
\noi The values of $\alpha$ show that for all indices, except I11 and I18, the proximity of the data to either LVLP or HVLP models, are more than 0.3. The LVLP semi-Markov binary regime switching MJD models fit significantly better to every index than its Markov counterpart (see Table \ref{Table2} for the comparison).

\subsection{Summary}
\begin{table}[ht]
	\centering
	\caption{Model fitting using $\alpha_4$-values from Table \ref{Table2} and Table \ref{Table3}} \label{Table4}	
\begin{tabular}{|l|cccc|l|l|}
		\hline
		\multicolumn{1}{|}{}& \multicolumn{4}{|c|}{$\alpha_4$-values}& \multicolumn{2}{|c|}{Model fitting}\\ \cline{2-7}
		\multicolumn{1}{|}{}& \multicolumn{2}{|c}{LVLP}& \multicolumn{2}{c|}{HVLP}&\multicolumn{1}{|l|}{Model class with}&\multicolumn{1}{|l|}{Estimated }\\
		{Index} &
		{M} &
		{SM} &    {M} &
		{SM} & {largest $\alpha_4$} & {$(\sigma(1), \sigma(2)) \ \ \frac{1}{\la_1\Delta}, \frac{1}{\la_2\Delta}, k$}\\
		\hline
  I01 &  0.03   &   0.20       &   \bf{0.40} &   \bf{0.40}& HVLP Markov         &\ (11.2\%,  6.2\%) \ 7,  40, 1       \\
  I02 &  0.06   &   0.27       &   \bf{0.38} &   \bf{0.38}& HVLP Markov         &\ (11.1\%,  6.3\%) \ 7,  40, 1       \\
  I03 &  0.02   &   0.21       &   \bf{0.41} &   \bf{0.41}& HVLP Markov         &\ (10.6\%,  6.2\%) \ 8,  45, 1       \\
  I04 &  0.05   &   0.25       &   \bf{0.46} &   \bf{0.46}& HVLP Markov         &\ (11.0\%,  5.9\%) \ 6,  34, 1       \\
  I05 &  0.09   &   0.31       &   \bf{0.34} &   \bf{0.34}& HVLP Markov         &\ (17.1\%,  8.5\%) \ 5,  28, 1       \\
  I06 &  0.16   &   \bf{0.38}  &   0.31      &   0.34     & LVLP semi-Markov    &\ (6.0\%, 10.1\%) 14,  79, 2       \\
  I07 &  0.21   &   \bf{0.41}  &   0.14      &   0.14     & LVLP semi-Markov    &\ (6.8\%, 10.8\%) 13,  74, 4       \\
  I08 &  0.15   &   \bf{0.39}  &   0.36      &   0.36     & LVLP semi-Markov    &\ (6.2\%,  9.7\%) \ 10,  57, 5       \\
  I09 &  0.26   &   0.39       &   \bf{0.42} &   \bf{0.42}& HVLP Markov         &\ (19.4\%,  9.6\%)  \ 5,  28, 1       \\
  I10 &  0.26   &   \bf{0.42}  &   0.30      &   0.30     & LVLP semi-Markov    &\ (6.6\%, 10.6\%) 16,  91, 2       \\
  I11 &  0.03   &   \bf{0.16}  &   0.07      &   0.07     & LVLP semi-Markov    &\ (6.5\%, 11.0\%) 15,  85, 2       \\
  I12 &  0.03   &   0.21       &   0.26      &   \bf{0.36}& HVLP semi-Markov    &\ (22.6\%, 12.1\%)  5,  28, 3       \\
  I13 &  0.39   &   \bf{0.43}  &   0.14      &   0.22     & LVLP semi-Markov    &\ (7.5\%, 16.5\%)  \ 6,  34, 2       \\
  I14 &  0.16   &   0.34       &   \bf{0.42} &   \bf{0.42}& HVLP Markov         &\ (16.4\%,  7.3\%)  \ 5,  28, 1       \\
  I15 &  0.12   &   \bf{0.31}  &   0.13      &   0.13     & LVLP semi-Markov    &\ (7.5\%, 11.7\%) 11,  62, 4       \\
  I16 &  0.07   &   0.34       &    0.35     &   \bf{0.43}& HVLP semi-Markov    &\ (13.6\%,  8.8\%)  \ 5,  28, 6       \\
  I17 &  0.33   &   0.38       &   \bf{0.39} &   \bf{0.39}& HVLP Markov         &\ (35.9\%, 15.0\%)  7,  40, 1       \\
  I18 &  0.05   &   0.22       &   0.27      &   \bf{0.29}& HVLP semi-Markov    &\ (13.6\%,  6.9\%)  \ 6,  34, 2       \\  \hline
\end{tabular}\end{table}
\noi In Table \ref{Table4} we summarize the comparison on fitting between all four different classes of models. In the last two columns, we record the best-fit model class and the best-fit parameter values. We do so by using the proximity measure $\alpha_4$ values obtained under each model class. The largest $\alpha_4$ values are highlighted with boldface in the table. The columns M and SM stand for the Markov and semi-Markov classes. The considered semi-Markov class subsumes the Markov class, i.e., we allow $k=1$ value. Except I11, the best-fit model for none of the indices has proximity less than $29\%$. In Table \ref{Table4} we observe that LVLP models with binary semi-Markov regimes fit strictly better than the Markov counter part to each index. A similar observation was made in \cite{DG19} which does not incorporate the jump discontinuities of asset price data. Since the class of semi-Markov(SM) regime models considered here subsumes the class of Markov(M) models, the $\alpha_4$ for semi-Markov class cannot be smaller than that of the Markov counter part. Therefore unless $\alpha_4$ for a semi-Markov class is strictly greater, we do not fit a semi-Markov model. It is important to note that, we have considered only a narrow class of semi-Markov models for illustration purpose and we still have obtained significantly better fit for some indices, including I10 and I15. A more detailed empirical study using a larger class of holding time distributions rather than only the gamma distribution, as considered here, may result in an improved fitting of semi-Markov models to most of the indices. It is needless to mention that the corresponding computational complexity would also increase significantly. For managing the computation time parallel algorithms have been implemented for measuring the proximity.

\section{Conclusion}\label{conclusion}
\noi In this paper, we have devised a test statistic, which is suitable for inference of the switching in diffusion term when the process has nontrivial jump discontinuities. More importantly, the jump detection is significantly accurate whereas the procedure is computationally inexpensive. This paper extends the scope of investigation significantly which was introduced in \cite{DG19}. In \cite{DG19}, only the continuous path regime switching models were considered for inference. Furthermore, the approach used in \cite{DG19} only works to infer the special case of binary regime where the low volatility regime occurs with low probability(LVLP). The approach adopted here is applicable for both types of binary regime switching models, namely low volatility with low probability or high volatility with low probability. Moreover, we have considered some historical data of Indian sectorial indices and performed the inference, developed in this paper. All Python codes, used in this paper, can be accessed from \href{https://github.com/SharanRajani/BinaryRegimeModelTesting}{https://github.com/SharanRajani/BinaryRegimeModelTesting}.

For inference of switching parameters we have maximized a proximity measure. In principle, maximisation of the proximity measure can be performed by various different algorithms. The algorithm that we have used is the direct method. In this method the computational complexity increases exponentially with the dimension of the domain. On the other hand that is linear for a gradient descent algorithm. The use of gradient descent becomes more relevant when the number of states is more than two. In future, with the help of gradient descent, we aim to develop inference of ternary regime switching models with the presence of jump discontinuities.

\section*{Acknowledgement}
Authors are grateful to Prof. Manjunath Krishnapur and Mr. D.V.S. Abhijit for some helpful discussion.

\section*{Funding Information}
\noindent The first author's research was supported by the Ministry of Science and Technology (MOST 108-2811-M-001-628, MOST 108-2118-M-001-003-MY2) of the Republic of China. The research of the second author was supported in part by the SERB MATRICS (MTR/2017/000543), DST FIST (SR/FST/MSI-105), NBHM 02011/1/2019/NBHM(RP)R\&D-II/585, and DST/INT/DAAD/P-12/2020. We also acknowledge National Supercomputing Mission (NSM) for providing computing resources of ‘PARAM Brahma’ at IISER Pune, which is implemented by C-DAC and supported by the Ministry of Electronics and Information Technology (MeitY) and Department of Science and Technology (DST), Government of India.

\section*{Conflict of Interest Statement}
The Authors declare that there is no conflict of interest.

\end{document}